\documentclass[journal,draftcls,onecolumn,12pt,twoside]{IEEEtran}

\normalsize

\usepackage{cite}
\usepackage{amsmath,amssymb,amsfonts}
\usepackage{algorithmic}
\usepackage{algorithm}
\usepackage{graphicx}
\usepackage{float}
\usepackage{footmisc}
\usepackage{stfloats}
\usepackage{textcomp}
\usepackage{multirow}
\usepackage{lscape}
\usepackage{caption}
\usepackage{subcaption}
\usepackage{enumerate}
\usepackage{xcolor}

\begin{document}
%
\title{Energy-Efficient Resource Allocation for Aggregated RF/VLC  Systems}

\author{Sylvester~Aboagye,~\IEEEmembership{Student~Member,~IEEE,}~Telex~M.~N.~Ngatched,~\IEEEmembership{Senior~Member,~IEEE,}~Octavia~A.~Dobre,~\IEEEmembership{Fellow,~IEEE,}~and~H.~Vincent~ Poor,~\IEEEmembership{Life~Fellow,~IEEE}



\thanks{This work has been submitted to the IEEE for possible publication. Copyright may be transferred without notice, after which this version may no longer be accessible.}
}

%



\maketitle
\vspace*{-19mm}

\begin{abstract}
\vspace*{-4mm}
Visible light communication (VLC) is envisioned as a core component of future wireless communication networks due to, among others, the huge unlicensed bandwidth it offers and the fact that it does not cause any interference to existing radio frequency (RF) communication systems. In order to take advantage of both RF and VLC, most research on their coexistence has focused on hybrid designs where data transmission to any user could originate from either an RF or a VLC access point (AP). However, hybrid RF/VLC systems fail to exploit the distinct transmission characteristics (e.g., susceptibility of VLC transmissions to blockages, limited field-of-view of VLC APs and receivers, more coverage and better reliability of RF systems, etc.) of RF and VLC systems to fully reap the benefits they can offer. Aggregated RF/VLC systems, in which any user can be served simultaneously by both RF and VLC APs, have recently emerged as a more promising and robust design for the coexistence of RF and VLC systems. To this end, this paper, for the first time, investigates AP assignment, subchannel allocation (SA), and transmit power allocation (PA)  to optimize the energy efficiency (EE) of aggregated RF/VLC systems while considering the effects of interference and VLC line-of-sight link blockages. A novel and challenging EE optimization problem is formulated for which an efficient joint solution based on alternating optimization is developed. More particularly, an energy-efficient AP assignment algorithm based on matching theory is proposed. Then, a low-complexity SA scheme that allocates subchannels to users based on their channel conditions is developed. Finally, an effective PA algorithm is presented by utilizing the quadratic transform approach and a multi-objective optimization framework. Extensive simulation results reveal that: 1) the proposed joint AP assignment, SA, and PA solution obtains significant EE, sum-rate, and outage performance gains with low complexity, and 2) the aggregated RF/VLC system provides considerable performance improvement compared to hybrid RF/VLC systems.
\end{abstract}

\vspace*{-10mm}
\begin{IEEEkeywords}
\vspace*{-4mm}
Visible light communication, radio frequency, aggregated systems, multi-objective optimization, energy efficiency, matching theory.
\end{IEEEkeywords}

\IEEEpeerreviewmaketitle

\section{Introduction}
\vspace*{-1mm}
\IEEEPARstart{V}{isible} light communication (VLC) has attracted significant research interest and is expected to be a key component of future communication systems \cite{8669813}. VLC, a communication technology that uses frequencies in the visible light spectrum, offers a vast amount of license-free bandwidth, high security due to the poor penetration of visible light signals, and relatively lower power consumption since the same power is used for the dual-purpose of illumination and communication. VLC uses readily available light sources such as light emitting diodes (LEDs) as transmitters and the receivers are equipped with photodetectors (PDs). Data transmission is achieved in VLC systems via intensity modulation at the transmitter side and direct detection at the receiver side.  

Like other high-frequency communication technologies (e.g., terahertz and millimeter-wave), signal propagation at such frequencies is short-range and highly susceptible to blockages. Hence, successful signal transmission in VLC requires a direct line-of-sight (LoS) path between the transmitter and the receiver. The inherent characteristics of visible light signals promote a mutually beneficial co-existence of VLC and radio frequency (RF) communication systems. More particularly, RF communication systems cannot support the ongoing rapid increase in demand for capacity due to the limited available radio spectrum in the sub-6 GHz band and the rising costs of installation and maintenance of RF cell sites. Hence, the co-existence of VLC and RF communication systems allows the combination of the former's high-speed data transmission and the latter's ubiquitous connectivity. It also provides a promising solution to the potential connectivity issues (resulting from blockages and users being in dead zones) in VLC.

The design of communication systems that allow the co-existence of VLC and RF systems can be realized in two main ways, namely, hybrid RF/VLC systems \cite{7593453,8013858,8949364,9484068,8966290,9159585,9411734,9578932,6965999,6655152,9148909,8110682,8741202,8395016,9667214,8862946,8970349} and aggregated RF/VLC systems \cite{7417378,7437374,7293077,7402263,9149834,9614989,8314706,8926487}. The former realizes signal transmission to any user via a VLC or an RF link, while the latter utilizes both VLC and RF links simultaneously to serve any user. However, the challenging problems of access point (AP) assignment due to the mixture of heterogeneous APs and the efficient allocation of transmit power and bandwidth resources arise in such communication systems \cite{8669813,8013858,9411734}. 
Specifically, the APs and receivers in VLC systems have limited field-of-view (FoV), affecting the strength of any received signal. As a result, the closest  AP might no longer provide the strongest channel gain \cite{9121765}. Moreover, network densification is expected to continue to play a vital role in the next generation of communication networks. This dense deployment can cause severe overlapping of the coverage areas, resulting in strong interference effects. Furthermore, the co-existence of RF/VLC systems will be characterized by multiple heterogeneous layers (e.g., macrocell, picocell, femtocell, and optical attocell layers) with different coverage sizes and operating characteristics \cite{7593453}. Hence, developing highly scalable and novel AP assignment and resource management schemes that can exploit the distinguishing characteristics of RF and VLC systems is of utmost importance. 

A number of studies have been carried out to tackle the AP assignment problem and/or the resource management issue in hybrid RF/VLC systems under various objectives such as sum-rate \cite{7593453,8013858,8949364,9484068,8966290,9159585,9411734,9578932}, spectral efficiency \cite{6965999,6655152,9148909}, power consumption \cite{8110682,8741202,8395016,9667214}, and energy efficiency (EE) \cite{8862946,9148909,8970349}. Unlike hybrid RF/VLC systems, few papers have considered the problem of resource allocation in aggregated RF/VLC systems and none has studied the AP assignment problem. The authors in \cite{8314706} investigated EE maximization via subchannel allocation (SA) and power allocation (PA) for an orthogonal frequency division multiple access (OFDMA)-based software-defined aggregated RF/VLC system with multiple RF and VLC APs. However, the authors made the simplifying assumption that the multiple LED arrays transmit the same signal simultaneously and, as a result, ignored any inter-cell interference (ICI) effects. In \cite{7417378}, the problem of transmit power optimization to maximize the achievable rate was investigated for an aggregated RF/VLC system with one VLC AP, one RF AP, and one user. 
In \cite{7437374}, the authors optimized the transmit power and bandwidth allocation to maximize the EE of an aggregated system, with a single RF AP and a single VLC AP, that serves multiple users. 
By leveraging the bonding technique in the Linux operating system, the design and real-time implementation of an aggregated system were explored in \cite{7293077,7402263}. The authors focused on a system with one RF AP, one VLC AP, and multiple users and provided a theoretical analysis of the average system delay.  
The authors in \cite{9149834} studied the EE maximization problem by optimizing the transmit power of an aggregated system with a single user, a single RF AP, and multiple VLC APs. In \cite{9614989}, the authors investigated the joint optimization of the discrete constellation input distribution and PA to maximize the achievable rate of an aggregated system with a single user, a single LED, and one RF antenna. The study in \cite{8926487} 
explored the PA optimization problem for an aggregated RF/VLC
system with a single RF AP, multiple VLC APs, and multiple users. However, the authors made the simplifying assumption that the coverage regions of the VLC APs do not overlap and, consequently, did not
consider ICI effects.

None of the studies mentioned above on aggregated RF/VLC systems (i.e., \cite{7417378,7437374,7293077,7402263,9149834,9614989,8314706,8926487}) consider the design and optimization of such a system with multiple APs and users with ICI for both the RF and VLC systems. Indoor environments are typically equipped with multiple LEDs with overlapping coverage areas to guarantee uniform illumination. Moreover, the overlapping illumination areas ensure seamless connectivity in VLC systems. The performance of aggregated RF/VLC systems with multiple APs (i.e., multiple RF and VLC APs) will suffer from ICI effects. Besides, the LoS blockage problem in VLC systems needs to be considered in optimizing the usage of any available resources and the performance analysis of aggregated RF/VLC systems. This is important since the proposed approaches in \cite{7417378,7437374,7293077,7402263,9149834,9614989,8314706,8926487} are not applicable to aggregated RF/VLC systems with multiple APs. Moreover, none of the above-mentioned papers considered the problems of AP assignment, SA, and transmit PA to optimize the EE of aggregated RF/VLC systems.

This paper investigates the EE optimization of aggregated RF/VLC systems equipped  with multiple RF and VLC APs serving multiple users, taking into account ICI effects and LoS blockages. The main contributions are summarized as follows:

\begin{itemize}
\item We consider an aggregated RF/VLC system composed of a single macrocell AP and multiple VLC and picocell APs, that serves multiple users. Under the electrical transmit power budgets for the APs and users' minimum quality-of-service (QoS) requirements, we study the joint optimization problem of AP assignment, SA, and transmit PA, while considering the effects of LoS blockages in the VLC systems and ICI in both communication systems. This aims to maximize the EE of the aggregated RF/VLC system. To the best of the authors' knowledge, this is the first time that such an EE optimization problem and constraint sets have been considered for an aggregated RF/VLC system.

\item The formulated design problem turns out to be a challenging non-convex optimization problem. To handle the non-convexity efficiently, the joint problem is decomposed into three subproblems, for which we propose a three-stage alternating solution technique. In the first stage, we exploit matching theory (MT) to assign users to APs while considering any ICI and blockage effects as well as the transmit power budgets of the APs. Then, in the second stage, each AP allocates its subchannels to the assigned users according to the quality of the channel condition. Finally, the APs optimize the transmit PA on the allocated subchannels such that the users' QoS requirements and the APs' transmit power budgets are satisfied while reducing any impact from blockages and ICI. For the transmit power optimization, the quadratic transform approach is first used to express the terms of the signal-to-interference plus noise ratio (SINR) into non-fractional forms. Then with a fixed AP assignment and SA,  the formulated EE optimization problem is recast as an equivalent multi-objective optimization problem (MOOP). We propose a solution for the MOOP based on the $\epsilon$-constraint method to obtain the globally optimal solution. 

\item Finally, we demonstrate the effectiveness of the proposed alternating solution for the joint problem and compare it with existing schemes and a hybrid RF/VLC system. Moreover, we also investigate the impact of LoS blockages and users' QoS requirements on the EE performance of the aggregated RF/VLC system.
\end{itemize}
The remainder of this paper is structured as follows. The considered system model as well as the channel models for the VLC and RF communication links are introduced and described in Section~\ref{lsm}. The proposed joint AP assignment, SA, and transmit PA optimization problem for aggregated RF/VLC systems is formulated and discussed in Section~\ref{leemp}. The proposed energy-efficient AP assignment and SA solutions are detailed in Section~\ref{leeaasa}, and the proposed solution for the PA subproblem is presented in Section~\ref{eepsca}. Section~\ref{srs} presents and analyzes the simulation results. Finally, Section~\ref{ccs} 
summarizes the work.
\vspace*{-3mm}
\section{System Model}\label{lsm}
\vspace*{-2mm}
\subsection{Aggregated RF/VLC Systems}
\begin{figure}
\centering
 \includegraphics[width=0.45\textwidth]{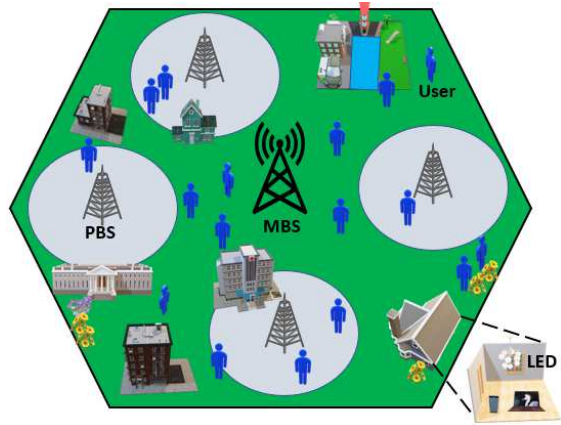}
 \vspace*{-4mm}
      \caption{Network model of a three--tier heterogeneous network.}
       \label{th}
       \vspace*{-1.1cm}
\end{figure}

\vspace*{-2mm}
Figure~\ref{th} illustrates the three-tier network model for the considered aggregated RF/VLC system. In this figure, the macro base station (MBS), also called a macrocell AP, provides blanket coverage for all users in the network. The pico base stations (PBSs), also called picocell APs,  provide smaller coverages such as hotspot areas, while the VLC APs are used exclusively for indoor data transmission. According to \cite{7593453}, such a network model involving the coexistence of RF and VLC systems provides several potential benefits that include: (i) high security induced by the poor penetration of the VLC signals; (ii) high total network capacity by employing picocell and VLC APs; (iii) high EE by realizing illumination and data transmission simultaneously in the VLC system; and (iv) reduced interference since the RF and VLC systems use different spectral bands. 

The set of RF and VLC APs are denoted by $\mathcal{K} = \left\{ {0, \ldots ,k \ldots ,\left|{\mathcal{K}}\right| - 1} \right\}$ and\\ $\mathcal{V} = \left\{ {1, \ldots ,v \ldots  ,\left|{\mathcal{V}}\right|} \right\}$, respectively, where the index $k=0$ represents the macrocell AP and $\left|{\cdot}\right|$ is the cardinality of a set. The RF and VLC APs employ the OFDMA scheme \cite{6825834,7360112}. The macrocell and the picocell APs use different sets of subchannels to avoid cross--tier ICI. However, the same subchannels are reused among all picocell APs and, as a result, there is co-tier ICI. The OFDMA subchannels for the macrocell and picocell APs are represented by the set $\mathcal{N}=\left\{1, \ldots,n, \ldots, \left|{\mathcal{N}}\right|\right\}$ and $\mathcal{M}=\left\{1, \ldots,m, \ldots, \left|{\mathcal{M}}\right|\right\}$, respectively. Each VLC AP in any indoor environment consists of an array of LEDs, and all attocells reuse the same set of subchannels. Hence, there is the occurrence of ICI in places where the illumination coverage of the VLC APs overlap. The VLC subchannels are represented by the set $\mathcal{Q}=\left\{1, \ldots,q, \ldots, \left|{\mathcal{Q}}\right|\right\}$. 

\begin{figure}
\centering
 \includegraphics[width=0.85\textwidth]{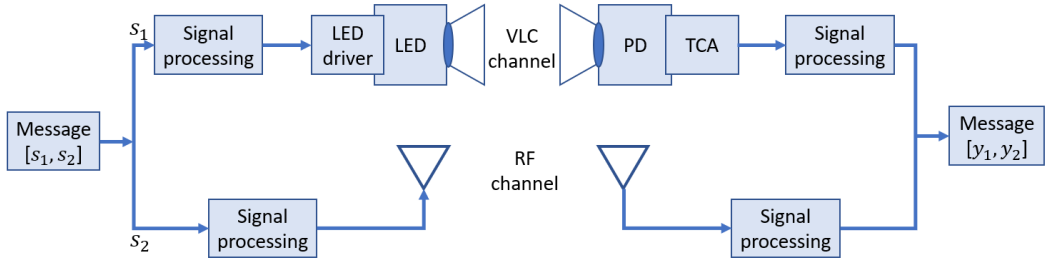}
  \vspace*{-4mm}
      \caption{Block diagram of data transmission in an aggregated RF/VLC system.}
       \label{mh}
       \vspace*{-8mm}
\end{figure}

The network serves $J$ users, represented by the set ${\mathcal J} = \left\{ {1, \ldots ,j, \ldots, \left|{\mathcal{J}}\right|} \right\}$, with multi-homing capability that allows any user to aggregate resources from RF and VLC APs, simultaneously. The users are uniformly and randomly distributed within the macrocell. A block diagram of signal transmission and reception in the downlink for any user with multi-homing capability is depicted in Fig.~\ref{mh}. In this figure, the message signal is transmitted simultaneously via the RF and VLC APs assigned to the user, where the signal $s_1$ is a real signal and $s_2$ is a complex signal for the VLC and RF links, respectively. The user's receiver comprises a single PD, with a transconductance amplifier (TCA) that converts the current output from the PD to voltage, and a single RF antenna for receiving the independently transmitted signal over the VLC and RF links, respectively. It is assumed in this work that the channel state information is known at the APs, and there are backhaul links between the macrocell AP and all the other APs for the reliable exchange of channel state information. The channel state information can be collected in the following way. Each AP broadcasts pilot signals to all users. Then, each user estimates the channel state information and sends it to the related AP via a feedback channel. Finally, all the APs send the channel state information to a centralized control unit (CCU).

\vspace*{-5mm}
\subsection{Channel Model}
\vspace*{-2.7mm}
\subsubsection{RF Channel} The channel power gain between  user $j$ and the macrocell AP on subchannel $n$ can be expressed as 

\vspace*{-5mm}
\begin{equation}\label{cg}
{G_{0,j}^n} = {10^{ - \frac{{{{L\left(d_{0,j}\right) + \Psi  + \Gamma  + {X_{\sigma}}}}\left[ {{\rm{dB}}} \right]}}{{10}}}},
\end{equation} 
where $L\left(\cdot\right)$ denotes the distance-dependent pathloss given by \cite{3gpp}:

\vspace*{-4mm}
\begin{equation}\label{pl1}
L\left(d_{0,j}\right)=128.1 + 37.6{\log _{10}}\left( {{d_{{{0,j}}}}} \right),
\end{equation}
with $d_{0,j}$ being the distance between user $j$ and the macrocell AP in km, $\Psi$ is the penetration loss which is defined as $\Psi=0\,{\rm dB}$ for outdoor users and  $\Psi=20\,{\rm dB}+0.5d$ for any indoor user, with $d$ being a distance parameter in m that takes an independent uniform random value from $\left[ {0,\min \left( {25,{d_{0,j}}} \right)} \right]$. The parameters $\Gamma$ and $X_{\sigma}$ represent the  multipath fading and the log-normal shadowing standard deviation, respectively. 

The channel power gain between user $j$ and picocell $k,\,k\ne0$ on  subchannel $m$ is 

\vspace*{-4mm}
\begin{equation}\label{cgp}
{G_{k,j}^m} = {10^{ - \frac{{{{L\left(d_{k,j}\right) + \Psi  + \Gamma  + {X_{\sigma}}}}\left[ {{\rm{dB}}} \right]}}{{10}}}},
\vspace*{-3mm}
\end{equation} 
where 

\vspace*{-4mm}
\begin{equation}\label{pl2}
L\left(d_{k,j}\right)=140.7 + 36.7{\log _{10}}\left( {{d_{{{k,j}}}}} \right),
\end{equation}
with $d_{k,j}$ being the distance between user $j$ and the AP $k$, and $\Psi=23\,{\rm dB}+0.5d$ for any indoor user served by any picocell AP with $d$ being a distance parameter with value from $\left[ {0,\min \left( {25,{d_{k,j}}} \right)} \right]$. 

\subsubsection{VLC Channel}
Only the LoS paths are considered as, according to \cite{7437374}, the non-LoS signals degrade significantly  and may result in unsuccessful data transmissions. The LoS channel power gain between user $j$ and the VLC AP $v$ on  subchannel $q$ can be expressed as follows:

\vspace*{-4mm}
\begin{equation}\label{cg1}
\begin{array}{*{20}{l}}
{{G_{v,j}^q} = \rho_{v,j}^q\frac{{{A_{{\rm{PD}}}}\left( {m_1 + 1} \right)}}{{2\pi d_{v,j}^2}}{\cos ^{m_1}}\left( {{\phi _{v,j}}} \right)T\left( {{\psi _{v,j}}} \right)}{G\left( {{\psi _{v,j}}} \right) {\cos }\left( {{\psi _{v,j}}} \right),}
\end{array}
\end{equation}
where $\rho_{v,j}^q$ is the probability of LoS availability (i.e., the probability that there is no obstacle in the communication link) between AP $v$ and user $j$ on subchannel $q$, $A_{\rm{PD}}$ is the physical area of the PD, $m_1$ is the order of the Lambertian emission which is calculated as $m_1 =  - {{{{\log }_2}\left( {\cos \left( {{\phi _{{1}/{2}}}} \right)} \right)}}^{-1}$, with ${{\phi _{{1}/{2}}}}$ as the LED's semi-angle at half power, ${\phi _{v,j}}$ represents the AP $v$ irradiance angle to user $j$, $\psi _{v,j}$ is the angle of incidence of AP $v$ to user $j$, $T\left( {{\psi _{v,j}}} \right)$ is the gain of the optical filter, and $G\left( {{\psi _{v,j}}} \right) = {{{f^2}}}/{{{{\sin }^2}{\psi _{{\rm{FoV}}}}}},\,0 \le {\psi _{v,j}} \le {\psi _{{\rm{FoV}}}}$, represents the gain of the non-imaging concentrator, where $f$ and ${\psi _{{\rm{FoV}}}}$ denote the refractive index and FoV, respectively. 
\vspace{-3mm}
\subsection{Achievable Rates}
Shannon's capacity formula for additive white Gaussian noise (AWGN) channels is used to represent the achievable data rate on any RF link for mathematical tractability in this work. Based on this equation, the achievable downlink rate on subchannel $n$ of the macrocell AP for user $j$ is calculated as 

\vspace{-3mm}
\begin{equation}\label{seq1}
    R_{0,j}^n=B_{\rm RF} \log_2 \left( 1 + \frac{p_{0,j}^n \left|G_{0,j}^m\right|^2}{N_{\rm RF}B_{\rm RF}}\right),
\end{equation}
where $B_{\rm RF}$ is the subchannel bandwidth, $p_{0,j}^n$ is the transmit power allocated to user $j$ on subchannel $n$, and $N_{\rm RF}$ is the power spectral density of AWGN at the RF receiver. Similarly, the downlink rate on subchannel $m$ of picocell AP $k$ for user $j$ is given by

\vspace{-3mm}
\begin{equation}\label{seq2}
R_{k,j}^m=B_{\rm RF} \log_2 \left( 1 + \frac{p_{k,j}^m \left|G_{k,j}^m\right|^2}{{\sum\limits_{j'\ne j} \sum\limits_{k'\ne {{k}}}  {p_{k',j'}^{{m}} { {{ \left|G_{k',j}^m\right|^2 }} } }   } +       N_{\rm RF}B_{\rm RF} }\right), k\ne 0,
\end{equation}
where $p_{k,j}^m$ is the transmit power from picocell AP $k$ to user $j$ on the subchannel $m$, and $j'$ and $k'$ denote other users and picocell APs that reuse the same subchannel $m$, respectively. 

In optical wireless communication systems in general, and VLC systems in particular, there is no suitable closed-form channel capacity formula. Thus, the following tight lower bound on the achievable data rate for user $j$ on  subchannel $q$ of the VLC AP $v$ is used \cite{6636053,9578932}

\vspace{-3mm}
\begin{equation}\label{ls}
R_{v,j}^q={ \rho_{v,j}^q B_{\rm VLC} \log_2 \left({ 1 + {\frac{\exp(1)}{2\pi}} \frac{p_{v,j}^{q} {\left( {{R_{\rm{PD}}}{G_{v,j}^{q}}} \right)}^2} {\sum\limits_{j'\ne j}  {\sum\limits_{v'\ne v}  p_{v',j'}^{q} {\left( {{R_{\rm{PD}}}{G_{v',j}^{q}}} \right)}^2 }   +  N_{\rm VLC}B_{\rm VLC}} }  \right)},
\end{equation}
where $B_{\rm VLC}$ is the subchannel bandwidth, $p_{v,j}^{q}$ is the electrical transmit power from VLC AP $v$ to user $j$ on the VLC subchannel $q$, $v'$ ranges over other VLC APs that reuse subchannel $q$ to serve other users, denoted as $j'$, and $N_{\rm VLC}$ is the power spectral density of AWGN at the PD.

According to the block diagram in Fig.~\ref{mh} for data transmission in aggregated RF/VLC system, the achievable data rate of user $j$ is given as 

\vspace*{-3mm}
\begin{equation}\label{adt}
    R_j=\sum\limits_{\forall v}\sum\limits_{\forall q} R_{v,j}^q + \sum\limits_{\forall k, k\ne0}\sum\limits_{\forall m} R_{k,j}^m +\sum\limits_{\forall n} R_{0,j}^n,
\end{equation}
and the sum of the achievable rates in the three-tier heterogeneous network is calculated as $R_T=\sum\limits_{\forall j} R_j$. The total transmit power allocated to user $j$ is given by

\vspace*{-3mm}
\begin{equation}\label{adt}
    P_j=\sum\limits_{\forall v}\sum\limits_{\forall q} p_{v,j}^q + \sum\limits_{\forall k, k\ne0}\sum\limits_{\forall m} p_{k,j}^m +\sum\limits_{\forall n} p_{0,j}^n,
\end{equation}
where the first, second, and third summation terms represent the total power consumed by the VLC, the picocell, and the macrocell APs, respectively.  The total transmit power used to serve all users in the entire network is calculated as 

\begin{equation}\label{tot}
P_T=P_{\rm PBS} \left( \left|{\mathcal{K}}\right| - 1\right) +P_{\rm MBS} + P_{\rm VLC}\left( \left|{\mathcal{V}}\right| \right) +\sum\limits_{\forall j} P_j,
\end{equation}
where $P_{\rm PBS}$, $P_{\rm MBS}$, and $P_{\rm VLC}$ denote the circuit power consumption for any picocell, macrocell, and VLC AP, respectively.

\section{Energy Efficiency (EE) Maximization Problem}\label{leemp}
The efficiency of any system is a measurable quantity determined by the ratio of its output to input. In the system model presented in Fig.~\ref{th}, efficiency can be seen as the extent to which the RF and VLC APs are assigned among the users,  the available subchannels are allocated to the users, and the available transmit power to the RF and VLC APs are utilized to provide users with at least their required data rates. To that end, the EE [in bit/Joule] can be defined as the ratio of the amount of data transmitted to the amount of power consumed in the network. The considered EE maximization problem via the joint optimization of AP assignment, SA, and transmit PA can be formulated as in (\ref{eep}). Under this formulation, the variables to be optimized are the AP assignment vector ${\bf x}$, the SA vector ${\bf s}$, the transmit PA vector ${\bf p}$, and the outage vector ${\bf a}$, where user $j$ is said to be in an outage (i.e., $a_j=0$) if that user is not assigned any subchannel  and $a_j=1$ means otherwise. Specifically, the AP assignment variables $x_{k,j}$ and $x_{v,j}$ denote the assignment of RF AP $k$ to user $j$ and that of VLC AP $v$ to user $j$, respectively. The SA variables $s_{k,j}^m$, $s_{0,j}^n$, and $s_{v,j}^q$ indicate the assignment of subchannel $m$ of picocell AP $k$ to user $j$,  subchannel $n$ of the macrocell AP to user $j$, and subchannel $q$ of VLC AP $v$ to user $j$, respectively. Similarly, $p_{k,j}^m$, $p_{0,j}^n$, and $p_{v,j}^q$ represent the transmit power allocated by picocell AP $k$ to user $j$ on  subchannel $m$, by the macrocell to user $j$ on subchannel $n$, and by VLC AP $v$ to user $j$ on subchannel $q$, respectively.  

\setlength{\abovedisplayskip}{-15pt}
\begin{equation}\label{eep}
\begin{array}{l}
\mathop {\max }\limits_{{\bf x},{\bf p},{\bf s},{\bf a}} \eta=\frac{R_T}{P_T}\\
{\rm{s}}{\rm{.t}}{\rm{.}}\\
C1: s_{k,j}^m \le x_{k,j},\,\,\forall k,j,m,k\ne 0,  \,\,\,\,\,\,\,\,\,\,\,\,\,\,\,\,\,\,\,\,\,\,\,\,\,\,\,\,\,\,\,\,\,\,\,\,C6: \sum\limits_{\forall k} x_{k,j}=1,\,\,\, \forall j,\\
\,\,\,\,\,\,\,\,\,\,\,\,s_{0,j}^n \le x_{0,j},\,\,\forall j,n,  \,\,\,\,\,\,\,\,\,\,\,\,\,\,\,\,\,\,\,\,\,\,\,\,\,\,\,\,\,\,\,\,\,\,\,\,\,\,\,\,\,\,\,\,\,\,\,\,\,\,\,\,\,\,\,\,\,\,\,\,\,C7: \sum\limits_{\forall v} x_{v,j}\le 1,\,\,\, \forall j,\\
\,\,\,\,\,\,\,\,\,\,\,\,s_{v,j}^q \le x_{v,j},\,\, \forall v,j,q,  \,\,\,\,\,\,\,\,\,\,\,\,\,\,\,\,\,\,\,\,\,\,\,\,\,\,\,\,\,\,\,\,\,\,\,\,\,\,\,\,\,\,\,\,\,\,\,\,\,\,\,\,\,\,\,\,C8:  {\rm R}_j\ge {\rm R}_{\min}a_j,\,\,\, \forall j,\\
C2: p_{k,j}^m \le s_{k,j}^m P_k,\,\, \forall k,j,m,k\ne0,  \,\,\,\,\,\,\,\,\,\,\,\,\,\,\,\,\,\,\,\,\,\,\,\,\,\,\,\,\,\, C9: p_{k,j}^m \ge 0, \,\, \forall k,j,m,k\ne0, \\
\,\,\,\,\,\,\,\,\,\,\,\,p_{0,j}^n \le s_{0,j}^n P_0,\,\,\forall j,n, \,\,\,\,\,\,\,\,\,\,\,\,\,\,\,\,\,\,\,\,\,\,\,\,\,\,\,\,\,\,\,\,\,\,\,\,\,\,\,\,\,\, \,\,\,\,\,\,\,\,\,\,\,\,\,\,\,\,\,\,\,\,\,\,\,\,\,\,   p_{0,j}^n \ge 0, \,\,\forall j,n, \\
\,\,\,\,\,\,\,\,\,\,\,\,p_{v,j}^q \le s_{v,j}P_v,\,\, \forall v,j,q,\,\,\,\,\,\,\,\,\,\,\,\,\,\,\,\,\,\,\,\,\,\,\,\,\,\,\,\,\,\,\,\,\,\,\,\,\,\,\,\,\,\, \,\,\,\,\,\,\,\,\,\,\,\,\,\,\,\,\,\,\,\,  p_{v,j}^q \ge 0, \,\, \forall v,j,q,\\
C3: \sum\limits_{\forall j}\sum\limits_{\forall m}p_{k,j}^m \le P_k,\forall k,k\ne 0,   \,\,\,\,\,\,\,\,\,\,\,\,\,\,\,\,\,\,\,\,\,\,\,\,\,\,\,\,\,\,\,\,\,\,\,\,  C10: s_{k,j}^m \in \{0,1\},\,\,\forall k,j,m,k\ne 0,\\
\,\,\,\,\,\,\,\,\,\,\,\,\sum\limits_{\forall j}\sum\limits_{\forall n}p_{0,j}^n \le P_0,\,\,\,\,\,\,\,\,\,\,\,\,\,\,\,\,\,\,\,\,\,\,\,\,\,\,\,\,\,\,\,\,\,\,\,\,\,\,\,\,\,\,\,\,\,\,\,\, \,\,\,\,\,\,\,\,\,\,\,\,\,\,\,\,\,\,\,\,\,\,\,\,\,\,\,\,\,\,\, s_{0,j}^n \in \{0,1\}, \,\, \forall j,n,\\
\,\,\,\,\,\,\,\,\,\,\,\,\sum\limits_{\forall j}\sum\limits_{\forall q}p_{v,j}^q \le P_v,\,\,\forall v,  \,\,\,\,\,\,\,\,\,\,\,\,\,\,\,\,\,\,\,\,\,\,\,\,\,\,\,\,\,\,\,\,\,\,\,\,\,\,\,\,\,\, \,\,\,\,\,\,\,\,\,\,\,\,\,\,\,\,\,\,\,\,\,\,\,\,\,    s_{v,j}^q \in \{0,1\}, \,\, \forall v,j,q,\\

C4: \sum\limits_{\forall j} s_{k,j}^m \le 1,\,\,\forall k,m,k\ne 0,  \,\,\,\,\,\,\,\,\,\,\,\,\,\,\,\,\,\,\,\,\,\,\,\,\,\,\,\,\,\,\,\,\,\,\,\,C11: x_{k,j} \in \{0,1\},\,\, x_{v,j} \in \{0,1\},\,\,\, \forall k,v,j,\\
\,\,\,\,\,\,\,\,\,\,\,\,\sum\limits_{\forall j} s_{0,j}^n \le 1,\,\,\forall n,  \,\,\,\,\,\,\,\,\,\,\,\,\,\,\,\,\,\,\,\,\,\,\,\,\,\,\,\,\,\,\,\,\,\,\,\,\,\,\,\,\,\,\,\,\,\,\,\,\,\,\,\,\,\,\,\,\,\,\,\,\,C12: a_{j} \in \{0,1\},\,\, \forall j.\\
\,\,\,\,\,\,\,\,\,\,\,\,\sum\limits_{\forall j} s_{v,j}^q \le 1,\,\,\forall v,q,\\
C5: a_j \left(\left|{\mathcal{N}}\right|+\left|{\mathcal{M}}\right|+\left|{\mathcal{Q}}\right|\right) \ge \sum\limits_{\forall k,k\ne0}\sum\limits_{\forall m}s_{k,j}^m + \sum\limits_{\forall n}s_{0,j}^n + \sum\limits_{\forall v}\sum\limits_{\forall q}s_{v,j}^q,\,\, \forall j,\\
\end{array}
\vspace*{-5mm}
\end{equation}

The physical meaning of the constraints in (\ref{eep}) is explained as follows. Constraint $C1$ ensures that any subchannel of an AP can only be allocated to a user if that user is assigned to that AP. For example, considering the picocell AP $k$ and user $j$, the variable $s_{k,j}^m$ can take the value of $0$ or $1$ when $x_{k,j}=1$, and can only take the value of $0$ when $x_{k,j}=0$. Constraint $C2$ implies that the transmit power on each subchannel should not exceed the maximum value specified by $P_k$, $P_0$, and $P_v$, for picocell AP $k$, the macrocell AP, and VLC AP $v$, respectively. Moreover, $C2$ ensures that no power is allocated to any user on any subchannel if that particular subchannel is not assigned to that user. Constraint $C3$ is the transmit power budget for the APs. Constraint $C4$ guarantees that any subchannel of an AP is allocated to at most one user. Constraint $C5$ ensures that no more subchannels than available are allocated to user $j$. Constraints $C6$ and $C7$ ensure that each user is assigned to one RF AP and at most to one VLC AP, respectively. Constraint $C8$ guarantees the minimum QoS requirement, $R_{\min}$. It requires that the aggregate data rate of  user $j$ not in an outage is constrained to be equal to or higher than $R_{\min}$. Constraints $C9$, $C10$, $C11$, and $C12$ are imposed to guarantee that the transmit power variables are non-negative, the SA variables are binary, the AP assignment variables are binary, and the outage variables are binary, respectively.

Problem (\ref{eep}) is unique for aggregated RF/VLC systems and has not been studied before. For instance, the problem requires that each user should be assigned to an RF AP and a VLC AP if that will impact the EE performance positively. However, a user assigned to an AP does not necessarily guarantee that any subchannel(s) will be allocated to that user from the AP as specified in constraint $C1$. Note that subchannels are scarce resources that should be utilized efficiently. Allocating them to any user when the channel condition is bad would require the AP to transmit at a higher transmit power to guarantee $R_{\min}$. This could result in high interference for users sharing the same subchannel. Moreover, the joint problem in (\ref{eep}) is difficult to solve directly due to the existence of both binary and continuous
variables, the non-convex SINR structure in the objective function, the coupling of the decision variables, the QoS requirement constraint, as well as the
fractional form of the objective function. This joint problem belongs to the class of mixed-integer nonlinear programming problems. To obtain the global optimal solution, a direct approach would involve an exhaustive search of all the possible AP assignment, SA, and PA combinations to find the solution that yields the highest EE performance. However, the computational complexity associated with the exhaustive search method is exponential and, as a result, infeasible in practice, even for small network sizes. Moreover, by decoupling this joint problem into three subproblems, each subproblem remains challenging to solve with conventional convex and quasi-convex optimization techniques. Specifically, the AP assignment and SA subproblems are combinatorial optimization problems, while the PA subproblem is challenging due to its non-convex structure. 

\begin{figure}
\centering
 \includegraphics[width=0.5\textwidth]{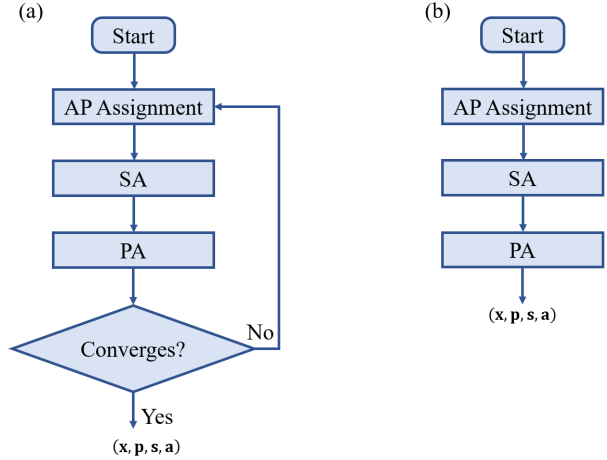}
      \caption{Framework to obtain the joint solution: (a) alternating optimization; (b) non-alternating.}
       \label{solf}
       \vspace*{-7mm}
\end{figure}

The proposed approach involves separately optimizing the AP assignment, the SA, and the transmit PA subproblems in order to reduce the associated computational complexity and solve the EE optimization problem in (\ref{eep}) faster \cite{8891923,8314706,9411734}. The joint solution to (\ref{eep}) can therefore be obtained by either alternating among the three subproblems until convergence or by just solving the three subproblems once but in a successive fashion, as illustrated in Fig.~\ref{solf}. Simulation results will be used to compare these two approaches.

\vspace*{-3mm}
\section{Energy-efficient AP Assignment and Subchannel Allocation (SA)}\label{leeaasa}
\vspace*{-2mm}
In  this section, the AP assignment and the SA optimization subproblems are investigated, and practical solution approaches are proposed. More particularly, the AP assignment subproblem is considered first, and a matching algorithm based on MT \cite{7105641,Roth1990}, is proposed to assign APs to users such that the EE performance is maximized. Then, each AP allocates the available subchannels to users according to the quality of the channel conditions.

\vspace*{-5mm}
\subsection{Energy-efficient Access Point (AP) Assignment}
The proposed matching algorithm for the AP assignment subproblem is described in this subsection for any given transmit PA. The main idea of this matching algorithm is explained as follows. Firstly, the potential EE performance for user $j$ within the coverage range of the RF AP $k$ and the VLC AP $v$ is calculated as in (\ref{x2}) and (\ref{ees1}), respectively.

\vspace*{-2mm}
\begin{equation}\label{x2}
{\rm EE}_{k,j} = \left\{ \begin{array}{l}
\frac{B_{\rm RF} \log_2 \left( 1 + \frac{{\overline P}_k \left|{\overline G}_{k,j}\right|^2}{N_{\rm RF}B_{\rm RF}}\right)}{P_{\rm MBS}+P_k},\,\,{\rm if}\,\, k=0,\\\\
\frac{B_{\rm RF} \log_2 \left( 1 + \frac{{\overline P}_k \left|{\overline G}_{k,j}\right|^2 }{{ \sum\limits_{k'\ne {{k}}}  {{\overline P}_{k'} { {{ \left|{\overline G}_{k',j}\right|^2 }} } }   } +       N_{\rm RF}B_{\rm RF} }\right)}{P_{\rm PBS}+P_k}, \,\,{\rm otherwise}.\\
\end{array} \right.
\end{equation}

\begin{equation}\label{ees1}
{\rm EE}_{v,j}=\frac{{ \overline{\rho}_{v,j} B_{\rm VLC} \log_2 \left({ 1 + {\frac{\exp(1)}{2\pi}} \frac{{\overline P}_v {\left( {{R_{\rm{PD}}}{\overline{G}_{v,j}}} \right)}^2} { {\sum\limits_{v'\ne v}  {\overline P}_{v'} {\left( {{R_{\rm{PD}}}{\overline{G}_{v',j}}} \right)}^2 }   +  N_{\rm VLC}B_{\rm VLC}} }  \right)}}    {P_{\rm VLC}+P_v}.
\end{equation}
In (\ref{x2}), ${\overline P}_k$, which is obtained by dividing the total power by the number of subchannels, is a predetermined transmit power that each user is allocated when assigned to RF AP $k$ and   ${\overline{G}_{k,j}}=\left(\sum\limits_{\forall m} G_{k,j}^m \right)/  \left|{\mathcal{M}}\right|$ is the average channel power gain between AP $k$ and user $j$ over all the AP's subchannels. In (\ref{ees1}), ${\overline P}_v$ is the predetermined transmit power that is allocated to any user that is assigned to VLC AP $v$, $\overline{\rho}_{v,j}=\left(\sum\limits_{\forall q} \rho_{v,j}^q \right)/  \left|{\mathcal{Q}}\right|$  is the average of the probability of LoS  availability between AP $v$ and user $j$, and ${\overline{G}_{v,j}}=\left(\sum\limits_{\forall m} G_{v,j}^q \right)/ \left|{\mathcal{Q}}\right|$ is the average channel power gain between AP $v$ and user $j$. Note that the EE definitions in (\ref{x2}) and (\ref{ees1}) accurately capture the LoS blockages for the VLC links as well as ICI effects in the VLC and RF communication systems. Secondly, each of RF AP $k$ and VLC AP $v$ constructs a preference list (PL) by sorting ${\rm EE}_{k,j}$ and  ${\rm EE}_{v,j}$ in decreasing order of $j$, respectively. Similarly, user $j$ builds a PL of the RF APs and VLC APs by sorting ${\rm EE}_{k,j}$ and  ${\rm EE}_{v,j}$ in decreasing order of $k$ and $v$, respectively. For the RF system, the PL of APs and users is denoted by the set ${\mathcal L}_{\rm RF}=\left\{ {{\bf l}_{{\rm RF}_0}^{\rm UT}, \ldots ,{\bf l}_{{\rm RF}_k}^{\rm UT} \ldots ,{\bf l}_{{\rm RF}_{\left|{\mathcal{K}}\right| - 1}} ^{\rm UT}, {\bf l}_1^{\rm RF}, \ldots, {\bf l}_j^{\rm RF}, \ldots,  {\bf l}_{\left|{\mathcal{J}}\right|}^{\rm RF} } \right\}$, where ${\bf l}_{{\rm RF}_k}^{\rm UT}$  denotes the preference relation of RF AP $k$ over the set of user terminals (UTs), while ${\bf l}_j^{\rm RF}$ represents the preference relation of user $j$ over the set of the available RF APs. Thus, the first user in ${\bf l}_{{\rm RF}_k}^{\rm UT}$ corresponds to $j^*=\arg \max_j {\rm EE}_{k,j},\,j\in {\mathcal{J}}$. Given RF AP $k$ and users $j_1,j_2\in {\mathcal{J}}$, it can be concluded that AP $k$ prefers $j_1$ to $j_2$ if $j_1$ precedes $j_2$ on AP $k$'s PL. For the VLC system, the PL of APs and users is denoted by the set ${\mathcal L}_{\rm VLC}=\left\{ {{\bf l}_{{\rm VLC}_1}^{\rm UT}, \ldots ,{\bf l}_{{\rm VLC}_v}^{\rm UT} \ldots ,{\bf l}_{{\rm VLC}_{\left|{\mathcal{V}}\right|}} ^{\rm UT}, {\bf l}_1^{\rm VLC}, \ldots, {\bf l}_j^{\rm VLC}, \ldots,  {\bf l}_{\left|{\mathcal{J}}\right|}^{\rm VLC} } \right\}$, where ${\bf l}_{{\rm VLC}_v}^{\rm UT}$ and ${\bf l}_j^{\rm VLC}$ represent the preference relation of VLC AP $v$ over the set of UTs and user $j$ over the available VLC APs, respectively. Finally, the energy-efficient AP assignment can therefore be formulated as a 4-tuple $\left({\mathcal{K}},{\mathcal{J}}, {\mathcal{P}}_{\rm RF}, {\mathcal{L}}_{\rm RF} \right)$ and $\left({\mathcal{V}},{\mathcal{J}}, {\mathcal{P}}_{\rm VLC}, {\mathcal{L}}_{\rm VLC} \right)$, for the RF and VLC systems, respectively, with  ${\mathcal{P}}_{\rm RF} = \left\{P_0,\ldots, P_k, \ldots, P_{{\mathcal K}-1} \right\}$  and ${\mathcal{P}}_{\rm VLC}=\left\{P_1,\ldots, P_v, \ldots, P_{\mathcal V}\right\}$ being the quotas for the RF and VLC APs, respectively, that indicate the available transmit power budget of each AP. It is desired to match the elements in the disjoint sets ${\mathcal{K}}$ and ${\mathcal{J}}$ for the RF system and ${\mathcal{V}}$ and ${\mathcal{J}}$ for the VLC system using the preference relations defined in ${\mathcal{L}}_{\rm RF}$ and ${\mathcal{L}}_{\rm VLC}$, respectively. Note that the AP assignment matching games for the RF and VLC systems can be implemented simultaneously and in parallel since our formulated matching game only considers the transmit power budgets.
A formal definition of this bilateral matching\footnote{The matching is bilateral because a user is associated with a  given AP if and only if that AP is assigned to that user.}  game is given as follows. 

A one-to-many matching $\mu_{RF}$ $\left(\mu_{VLC}\right)$ is defined as a mapping from the set ${\mathcal K} \cup {\mathcal J}$ $\left({\mathcal V} \cup {\mathcal J}  \right)$ into the set of all subsets of ${\mathcal K} \cup {\mathcal J}$ $\left({\mathcal V} \cup {\mathcal J}  \right)$ such that for each $k\in {\mathcal K}$, $v\in {\mathcal V}$ and $j\in {\mathcal J}$:

\begin{enumerate}[(a)]
\item $\left|\mu_{\rm RF} \left( j \right) \right| = 1$ for every user $j$, where $\mu_{\rm RF} \left( j \right)=k$ denotes that user $j$ is assigned to RF AP $k$ at the matching $\mu_{\rm RF}$, and $\left|\mu_{\rm VLC} \left( j \right) \right| \le 1$ for every user $j$, where $\mu_{\rm VLC} \left( j \right)=v$ indicates that user $j$ is assigned to VLC AP $v$ at the matching $\mu_{\rm VLC}$.
\item $\left|\mu_{\rm RF} \left( k \right) \right| {\overline P}_k \le  P_k$ for every RF AP $k$ and $\left|\mu_{\rm VLC} \left( v \right) \right| {\overline P}_v \le  P_v$ for every VLC AP $v$.
\item $\mu_{\rm RF} \left( j \right)=k$ if and only if $\mu_{\rm RF} \left( k \right)=j$, and $\mu_{\rm VLC} \left( j \right)=v$ if and only if $\mu_{\rm VLC} \left( v \right)=j$.
\end{enumerate}
Condition (a) ensures that each user is matched to only one RF AP and at most to one VLC AP. Condition (b) guarantees that the total power allocated to the matched users by any RF and VLC APs does not exceed the available power budget (i.e., quota). Condition (c) states that if a user $j$ is matched to the RF AP $k$, this AP $k$ is also matched to the same user $j$, and the same can be said about the VLC APs. The QoS requirement defined in constraint $C8$ of (\ref{eep}) is not included in the AP matching game since the existence of $R_{\min}$ will severely restrict the matching of users and APs and, as a result, affect the quality of the solution obtained from the matching game. The QoS requirement is considered in the PA subproblem, where the users' $R_{\min}$ can be satisfied by adjusting the PA coefficients.   

The proposed algorithm to solve this bilateral matching game and obtain the global optimal solution to the EE AP assignment subproblem is summarized in Algorithm~\ref{A1}. This matching procedure is assumed to be performed by a CCU located at the macrocell AP or in the cloud and provided with any required input data. In this algorithm, the CCU takes as input data the initial transmit power values, the average channel gain information, the LoS availability information, the PLs of both users and APs, and the quota of the APs, and delivers a final matching relation $\mu^*$. In the initialization stage, the CCU denotes the preference index of  user $j$  as $t_j$ with $t_j=1, \forall j$ and also represents the waitlist of RF AP $k$ as ${\mathcal W}_k=\emptyset$ and VLC AP $v$ as ${\mathcal W}_v=\emptyset$. At the $t_j$-th iteration of the matching game, user $j$ proposes to match with its top-ranked AP and removes this AP from its PL (thus, this is done in parallel for the RF and VLC system). RF AP $k$ (VLC AP $v$) places on its waitlist ${\mathcal W}_k$ $\left({\mathcal W}_v \right)$ the top-ranked users such that the sum of their allocated transmit powers do not exceed RF AP $k$'s (VLC AP $v$'s) quota $P_k$ ($P_v$). Thus, the PL's size of any user reduces by one at the end of each iteration, and the second-ranked AP at the $t_j$-th iteration becomes the first ranked one at the start of the $t_j+1$-th iteration. During the $t_j+1$-th iteration, each user submits a proposal to its most preferred RF and VLC APs on their respective updated PLs. Once again, each AP selects the top-ranked user among the new applicants and those on its waitlist, then places the selected user on an updated waitlist while rejecting the rest. This matching procedure terminates when every user is either on a waitlist (i.e.,  ${\mathcal W}_k\ne\emptyset, \,k\in {\mathcal K}$ and ${\mathcal W}_v\ne\emptyset,\,v\in{\mathcal V}$) or has been rejected by every AP on its RF and VLC PLs (i.e., $l_j^{\rm RF}=\emptyset,\,l_j^{\rm VLC}=\emptyset,\,j\in{\mathcal J}$). At this point, each AP accepts the users on its waitlist, and a final stable matching ${\mu}^*$  has been obtained. The AP for all users can be computed from the final matching according to 

\vspace*{-3mm}
\begin{equation}\label{msol}
    {x_{k,j}^*} = \left\{ {\begin{array}{*{20}{l}} 
{1,\,\, {\rm if}\, {\mu^* \left(j\right)}=k, \forall k\in{\mathcal K} }
\\
{0{\mkern 1mu} ,{\kern 1pt} {\mkern 1mu} {\kern 1pt} {\mkern 1mu} {\kern 1pt} {\mkern 1mu} {\kern 1pt} {\rm{otherwise}}{\rm{,}}}
\end{array}} \right. {\rm and}\,\,\,\, {x_{v,j}^*} = \left\{ {\begin{array}{*{20}{l}} 
{1,\,\, {\rm if}\, {\mu^* \left(j\right)}=v, \forall v\in{\mathcal V} }
\\
{0{\mkern 1mu} ,{\kern 1pt} {\mkern 1mu} {\kern 1pt} {\mkern 1mu} {\kern 1pt} {\mkern 1mu} {\kern 1pt} {\rm{otherwise}}{\rm{,}}}
\end{array}} \right.
\end{equation}
for RF AP $k$ and VLC AP $v$, respectively.

\begin{algorithm}
\caption{Energy-efficient MT-based AP Assignment.}
\label{A1}
\begin{algorithmic}
\STATE {{\textbf{Input:}} ${\overline P}_k$, ${\overline G}_{k,j}$, $\overline{\rho}_{v,j}$, ${\overline P}_v$, $\overline{G}_{v,j}$, ${\mathcal L}_{\rm RF}$, ${\mathcal L}_{\rm VLC}$, ${\mathcal P}_{\rm RF}$, and ${\mathcal P}_{\rm VLC}$, with $k\in {\mathcal K}$, $j\in{\mathcal J}$, and $v\in{\mathcal V}$.}

\STATE {{\textbf{Initialization:}} Set iteration counter for user $j$ as $t_j=1$,  and let the waitlists for the APs be denoted by ${\mathcal W}_k=\emptyset$ for RF AP $k$ and ${\mathcal W}_v=\emptyset$ for VLC AP $v$, with $k\in {\mathcal K}$, $j\in{\mathcal J}$, and $v\in{\mathcal V}$.}
\WHILE{$l_j^{\rm RF}\ne \emptyset$, $\forall j\in {\mathcal J}$ and ${\mathcal W}_k=\emptyset$, $\forall k\in {\mathcal K}$}

\STATE {(i) At the $t_j$-th iteration, user $j$ sends a proposal request to the $t_j$-th preferred RF AP in its PL (i.e., $l_j^{\rm RF}$) and clears that AP from the PL.}

\STATE {(ii) For each of the RF APs, AP $k$ considers all the proposal requests from the users and places on the waitlist ${\mathcal W}_k$ the highest-ranked users in its PL and rejects proposals when the quota $P_k$ is reached.}

\STATE {(iii) Set $t_j=t_j+1$.}

\ENDWHILE

\WHILE{$l_j^{\rm VLC}\ne \emptyset$, $\forall j\in {\mathcal J}$ and ${\mathcal W}_v=\emptyset$, $\forall v\in {\mathcal V}$}

\STATE {(i) At the $t_j$-th iteration, user $j$ sends a proposal request to the $t_j$-th preferred VLC AP in its PL (i.e., $l_j^{\rm VLC}$) and clears that AP from the PL.}

\STATE {(ii) For each of the VLC APs, AP $v$ considers all the proposal requests from the users and places on the waitlist ${\mathcal W}_v$ the highest-ranked users in its PL and rejects proposals when the quota $P_v$ is reached.}

\STATE {(iii) Set $t_j=t_j+1$.}

\ENDWHILE

\STATE {{\textbf {Output:}} The APs accept all users on the waitlists to form the stable matching $\mu^*$ and the AP assignment solution ${\bf x}^*$ can be computed from (\ref{msol}).
}
\end{algorithmic}
\end{algorithm}
\setlength{\textfloatsep}{29pt}

\vspace*{-7mm}
\subsection{Analysis of Stability, Optimality, and Convergence} 
\vspace*{-1mm}
In this subsection, the properties of the proposed matching algorithm for the energy-efficient AP assignment subproblem  are analyzed. Before discussing the stability property, the definition of a {\textit{blocking pair}} is provided. 

\textit{Definition 1:} 
Any pair of user $j\in {\mathcal J}$ and RF AP $k\in{\mathcal K}$ or user $j\in {\mathcal J}$ and VLC AP $v\in{\mathcal V}$ is said to be a blocking pair if user $j$ and the RF AP $k$ or user $j$ and the VLC AP $v$ prefer each other over their partners in the current matching.

\textit{Definition 2:} A matching is stable if there is no blocking pair. 

The above definition of stability implies that there is no pair of user and AP or an unhappy user or an unhappy AP that prefers being matched to each other or to another AP or to another user instead of being matched to their current partner.

The proposed matching algorithm in Algorithm~\ref{A1} is guaranteed to converge to a stable matching $\mu_{\rm RF}^*$ and $\mu_{\rm VLC}^*$ for the RF and VLC systems, respectively, for any stated preferences. The reason is that, at the end of Algorithm~\ref{A1}, user $j^*$ is matched with the top-ranked (i.e., most preferred) RF AP $k^*$ on its final updated PL, $l_{j^*}^{\rm RF}$, under the matching $\mu_{\rm RF} \left( j^* \right)$. For the VLC system, user $j^*$ is matched with the top-ranked AP $v^*$ on its final updated PL, $l_{j^*}^{\rm VLC}$, under the matching $\mu_{\rm VLC} \left( j^* \right)$. This matching is stable since RF AP $k$ (VLC AP $v$)  that user $j^*$ originally ranked higher than $k^*$ ($v^*$) was deleted from the PL $l_{j^*}^{\rm RF}$ $\left(l_{j^*}^{\rm RF}\right)$ after user $j^*$ sent a proposal request and got rejected. Therefore, the final matching gives RF AP $k$ and VLC AP $v$ a user that it ranked higher than $j^*$. It can be concluded that Algorithm~\ref{A1} produces the stable matching $\mu_{\rm RF}^*$ and $\mu_{\rm VLC}^*$ which is not blocked by RF AP $k$-user $j$ pair and VLC AP $v$-user $j$ pair, respectively.

Algorithm~\ref{A1} is guaranteed to produce a matching that gives each user $j^*$ its highest ranked VLC AP $v^*$ and/or its highest ranked RF AP $k^*$ and, as a result, obtains the globally optimal AP assignment solution. This follows from the fact that the output of  Algorithm~\ref{A1} is a stable matching. Thus, the proposed algorithm guarantees that the final matching gives users the APs that contribute to the highest network EE and only rejects the proposals of users that cannot be accepted by APs in any stable matching.

Finally, Algorithm~\ref{A1} is guaranteed to converge in a finite number of iterations, which is upper bounded by the number of APs, since no user sends more than one proposal request to any AP.

\vspace*{-5mm}
\subsection{Subchannel Allocation (SA) Scheme}
\vspace*{-1mm}
Having obtained the AP assignment solution from Algorithm~\ref{A1}, a low-complexity sub-optimal SA is proposed in this subsection. This scheme assigns any subchannel of an AP to a user based on the quality of the channel condition. A sub-optimal scheme is motivated because the optimal SA scheme, i.e., the exhaustive search, needs to search all possible combinations of users and subchannels for all APs and select the solution that maximizes the EE of the aggregated system. However, the task of enumerating all the candidate SA solutions dramatically increases the associated complexity of the exhaustive search. In comparison, the proposed SA scheme is more straightforward and can tackle the SA subproblem faster. 

The main idea of the SA scheme is that the macrocell AP, any picocell AP $k\in{\mathcal K},k\ne0$, and any VLC AP $v\in{\mathcal V}$ should allocate any subchannel $n\in{\mathcal N}$, $m\in{\mathcal M}$, and  $q\in{\mathcal Q}$, respectively, to the user with the highest channel power gain. For instance, the macrocell AP assigns subchannel $n$ to user $j^*$ (i.e., ${s_{0,j^*}^n}=1$) if $j^*=\mathop{\arg\max}_{j} {  {G_{0,j}^n} }$. Similarly,  picocell $k$ and VLC AP $v$ perform SA according to ${s_{k,j^*}^m}=1$ if  $j^*=\mathop{\arg\max}_{j} {  {G_{k,j}^m} }$ and ${s_{v,j^*}^q}=1$ if $j^*=\mathop{\arg\max}_{j} {  {G_{v,j}^q} }$, respectively. Based on the SA solution ${\bf s}^*$, any user $j\in{\mathcal J}$ can be said to be in outage (i.e., $a_j=0$) if $\sum\limits_{n\in{\mathcal{N}}} s_{0,j}^n + \sum\limits_{m\in{\mathcal{M}}} s_{k,j}^m +\sum\limits_{q\in{\mathcal{Q}}} s_{v,j}^q =0$. Such user can later try to access the network via the network's admission control scheme.

\vspace*{-5mm}
\section{Energy-efficient PA Scheme: $\epsilon$-Constraint Approach}\label{eepsca}
\vspace*{-1mm}
Given the AP assignment and SA solutions, the transmit PA is optimized in this section to maximize the EE of the aggregated RF/VLC system. Specifically, the energy-efficient PA subproblem can be formulated as 

\vspace*{-2mm}
\begin{equation}\label{eepa}
\begin{array}{l}
\mathop {\max }\limits_{\bf p} \eta=\frac{R_T}{P_T}\\
{\rm{s}}{\rm{.t}}{\rm{.}}\\
C2,\,C3,\,C8,\,{\rm and}\,\, C9.

\end{array}
\end{equation}
The PA subproblem above is non-convex since (i) the objective function is in a fractional form with respect to $\bf p$ and (ii) there are ICI terms in the rate function of the objective function and the QoS requirement in $C8$. Moreover, problem (\ref{eepa}) can be classified as a MOOP and is hard to solve in general since it involves two conflicting objectives, namely, maximizing the sum-rate while minimizing the total power consumption. Typically, there is no single global solution; rather, there is a set of acceptable trade-off optimal solutions called the {\textit{Pareto optimal set}} and corresponding objective function values called the {\textit{Pareto optimal frontier}}. A solution belongs to this set if no other solution can improve one of the objective functions without reducing the other objective function values.
Although the MOOP in (\ref{eepa}) can be converted into a single objective function (i.e., by rewriting the objective function into a parametric subtractive form) and then tackled by the well known Dinkelbach algorithm \cite{dink}, such an approach has several limitations including \cite{1599245}: 1) the objective function in a parametric subtractive form leads to only one solution and system engineers may desire to know all possible optimization solutions; 2) trade-offs between the objectives (i.e., sum-rate and total power) cannot be easily evaluated; and 3) the solution may not be attainable unless the search space is convex.  

In this section, a low-complexity solution, based on the framework of the $\epsilon$-constraint method for  MOOPs \cite{chir}, is proposed to discover the entire Pareto optimal frontier of (\ref{eepa})  that also contain the global optimal solution. The  concept of {{\textit {Pareto dominance}}} is first introduced. 

\textit{Definition 3 (Pareto dominance):} Given two solution vectors ${\bf p}^{\left(1\right)}$ and ${\bf p}^{\left(2\right)}$, ${\bf p}^{\left(1\right)}$ is said to {\textit {Pareto dominate}} ${\bf p}^{\left(2\right)}$, if and only if (i) solution ${\bf p}^{\left(1\right)}$ is no worse than ${\bf p}^{\left(2\right)}$ in all objectives, and (ii) solution ${\bf p}^{\left(1\right)}$ is strictly better than ${\bf p}^{\left(2\right)}$ in at least one objective. Thus, solution ${\bf p}^{\left(1\right)}$ is non-dominated by ${\bf p}^{\left(2\right)}$.

According to the $\epsilon$-constraint method, the EE optimization problem in (\ref{eepa}) can be cast as 

\vspace{-6mm}
\setlength{\abovedisplayskip}{0pt}
\begin{equation}\label{eepab}
\begin{array}{l}
\mathop {\min }\limits_{\bf p} {P_T}\\
{\rm{s}}{\rm{.t}}{\rm{.}}\\
C2,\,C3,\,C8,\,C9,\\
C13: R_T\ge {\epsilon},\,\,
\end{array}
\end{equation}
where $\epsilon=\lambda\times R_{\max}$ with $\lambda\in (0,1]$ and $R_{\max}$ determined from

\vspace{-6mm}
\begin{equation}\label{eepac}
\begin{array}{l}
R_{\max}=\mathop {\max }\limits_{\bf p} {R_T}\\
{\rm{s}}{\rm{.t}}{\rm{.}}\\
C2,\,C3,\,C8,\,{\rm and}\,C9.
\end{array}
\end{equation}
In (\ref{eepab}), the numerator function (i.e., the sum-rate) of the original EE problem in (\ref{eepa}) has been transformed into the constraint $C13$ that requires that the sum-rate of the aggregated RF/VLC system, $R_T$, should be greater than or equal to $\epsilon$. Thus, $\epsilon$ represents a lower bound of the value of $R_T$. The motivation for moving the sum-rate term to the constraint set is because the achievable rate is a function of the transmit power and, as a result, the impact of the total power consumed on the EE is much more significant than that of the sum-rate. By choosing different values for $\lambda$ and repeatedly solving (\ref{eepab}), we can generate its complete Pareto optimal set \cite{chir}. Specifically, for any value of $\lambda$ (and $\epsilon$), the resulting problem with $C13$ divides the original feasible objective space into two portions, $R_T\ge \epsilon$ and $R_T< \epsilon$. The right portion becomes the feasible solution of the resulting problem stated in (\ref{eepab}). In this way, intermediate Pareto optimal solutions can be obtained for nonconvex objective space problems as the unique solution of the  $\epsilon$-constraint problem stated in (\ref{eepab}) is Pareto optimal for any given lower bound $\epsilon$. To solve (\ref{eepab}), the value of $R_{\max}$ must be determined  first by solving (\ref{eepac}), which is a sum-rate optimization problem. An approach for solving the non-convex problem in  (\ref{eepac}) is proposed below.

\subsection{Determining $R_{\max}$}
Problem (\ref{eepac}) is highly intractable and non-convex because of the SINR terms in both the objective function and the constraint $C8$. To overcome this difficulty, the quadratic transform approach, originally proposed in \cite{fp}, is used to transform the fractional SINR terms into an equivalent non-fractional form. According to the quadratic transform technique, any SINR term in the RF system can be equivalently represented as 

\vspace{-5mm}
\begin{equation}\label{qt1}
\begin{array}{l}
{\rm SINR}_{k,j}^m= 2y_{k,j}^m \sqrt{{p_{k,j}^m \left|G_{k,j}^m\right|^2}} -y_{k,j}^m \left({{\sum\limits_{j'\ne j} \sum\limits_{k'\ne {{k}}}  {p_{k',j'}^{{m}} {{{ \left|G_{k',j}^m\right|^2 }} } }   } +       N_{\rm RF}B_{\rm RF} } \right),
\end{array}
\end{equation}
where $y_{k,j}^m$ is an auxiliary variable for the SINR term introduced by the application of the quadratic transform technique. For the VLC system, a similar transformation can be carried out according to

\vspace{-5mm}
\begin{equation}\label{qt2}
\begin{array}{l}
{\rm SINR}_{v,j}^q= 2y_{v,j}^q \sqrt{p_{v,j}^{q} {\left( {{R_{\rm{PD}}}{G_{v,j}^{q}}} \right)}^2} -y_{v,j}^q \left({\sum\limits_{j'\ne j}  {\sum\limits_{v'\ne v}  p_{v',j'}^{q} {\left( {{R_{\rm{PD}}}{G_{v',j}^{q}}} \right)}^2 }   +  N_{\rm VLC}B_{\rm VLC}} \right).
\end{array}
\end{equation}
By utilizing the SINR terms in (\ref{qt1}) and (\ref{qt2}) to calculate the achievable rate of any user as well as the sum-rate $R_T$, problem (\ref{eepac}) can be equivalently reformulated as 

\vspace*{-4mm}
\begin{equation}\label{abn1}
\begin{array}{l}
R_{\max}=\mathop {\max }\limits_{{\bf p},{\bf y}} {R_T}\\
{\rm{s}}{\rm{.t}}{\rm{.}}\\
C2,\,C3,\,C8,\,{\rm and}\,C9.
\end{array}
\end{equation}
Although, (\ref{abn1}) remains non-convex in ${\bf p}$ and ${\bf y}$, it becomes a convex optimization problem when ${\bf y}$ is fixed and the optimal solution can be obtained using the CVX toolbox \cite{cvx}. For a fixed ${\bf p}$, the optimal solution for ${\bf y}$, denoted by $\hat{\bf y}$ can be obtained in closed form by solving 
${\partial {{\rm SINR}_{k,j}^m}}/{\partial y_{k,j}^m}=0$ for the RF system and ${\partial {{\rm SINR}_{v,j}^q}}/{\partial y_{v,j}^q}=0$ for the VLC system. Specifically,

\begin{equation}\label{edas1}
\begin{array}{l}
{\hat y}_{k,j}^n=\frac{\sqrt{{p_{k,j}^m \left|G_{k,j}^m\right|^2}}} {{\sum\limits_{j'\ne j} \sum\limits_{k'\ne {{k}}}  {p_{k',j'}^{{m}} {\left( {{ \left|G_{k',j}^m\right|^2 }} \right)} }   } +       N_{\rm RF}B_{\rm RF} },\,\,\,\,\,\,\,\,\,\,\,
{\hat y}_{v,j}^q=\frac{\sqrt{p_{v,j}^{q} {\left( {{R_{\rm{PD}}}{G_{v,j}^{q}}} \right)}^2}}  {{\sum\limits_{j'\ne j}  {\sum\limits_{v'\ne v}  p_{v',j'}^{q} {\left( {{R_{\rm{PD}}}{G_{v',j}^{q}}} \right)}^2 }   +  N_{\rm VLC}B_{\rm VLC}}}.
\end{array}
\end{equation}
Then, the optimal ${p}$ for any fixed ${\bf y}$ can be obtained by solving the resulting convex problem in (\ref{abn1}). The proposed algorithm to determine the value for $R_{\max}$ is summarized in Algorithm~\ref{A3}.

\begin{algorithm}
\caption{Proposed Algorithm to Determine $R_{\max}$}
 \label{A3}
 \begin{algorithmic}
  \STATE {Set the iteration counter $c=1$, maximum error tolerance $\varepsilon > 0$, $R_{\max}^{(c)} > \varepsilon$, $R_{\max}^{(0)} =  0$, and initialize ${\bf p}^{(c)}$ using the equal power assignment (EPA) scheme, where $p_{0,j}^n=\frac{P_0}{\left|{\mathcal{N}}\right|}$,\\ $p_{k,j}^m=\frac{P_k}{\left|{\mathcal{M}}\right|},\,\forall k,k\ne0,$ and $p_{v,j}^q=\frac{P_v}{\left|{\mathcal{Q}}\right|},\,\forall v$; }
 \WHILE {$R_{\max}^{(c)} - R_{\max}^{(c-1)} > \varepsilon$}
        \STATE {$c = c + 1$;}
        \STATE {Calculate ${\bf y}^{(c)}$ using ${\bf p}^{(c-1)}$ and (\ref{edas1});}
        \STATE {Solve the convex problem (\ref{abn1}) given ${\bf y}^{(c)}$ to obtain ${R_{\max}^{(c)}}$ and ${\bf p}^{(c)}$;}
 \ENDWHILE
 \STATE {{\textbf{Output:}} ${R_{\max}}$}.
 \end{algorithmic}
\end{algorithm}
\setlength{\textfloatsep}{0pt}
\vspace*{-4mm}
\subsection{Determining the EE Solution}
Given the value of $R_{\max}$ and any value for ${\lambda}$, the transmit power minimization problem in (\ref{eepab}) can be formulated as in (\ref{abt}) after replacing the SINR terms with their equivalent quadratic forms given in (\ref{qt1}) and (\ref{qt2}).

\begin{equation}\label{abt}
\begin{array}{l}
\mathop {\min }\limits_{{\bf p},{\bf y}} {P_T}\\
{\rm{s}}{\rm{.t}}{\rm{.}}\\
C2,\,C3,\,C8,\,C9,\,{\rm and}\,C13.
\end{array}
\end{equation}
 In (\ref{abt}), the decision variables are the transmit power vector ${\bf p}$ and the auxiliary variable ${\bf y}$. It is non-convex in both ${\bf p}$ and ${\bf y}$
 due to the constraints in $C8$ and $C13$. However, for a fixed ${\bf y}$, (\ref{abt}) becomes a convex optimization problem. On the other hand, the optimal solution for ${\bf y}$ can be determined using the closed form expressions in (\ref{edas1}). Thus, for any given value of $\lambda$ (and its corresponding $\epsilon$ value), problem (\ref{abt}) is solved by optimizing  ${\bf p}$ and ${\bf y}$ in an alternating fashion until convergence. The proposed algorithm for solving (\ref{abt}) and obtaining the optimal EE solution, denoted by $\eta^*$, is summarized in Algorithm~\ref{A4}.  

\begin{algorithm}
\caption{Proposed Algorithm to Solve ({\ref{eepa}})}
 \label{A4}
 \begin{algorithmic}
 \STATE {{\textbf{Input:}} ${R_{\max}}$;}
 \STATE {Set $\lambda=0$, step size $\mu=0.1$, outer iteration counter $t=1$, maximum error tolerance $\varepsilon > 0$, and $\gamma^{(0)} =  0$; }
 \STATE {Create an empty non-dominated set ${\boldsymbol {\mathcal G}}$ and Pareto optimal front ${\boldsymbol \eta}$;}
 \WHILE {$\lambda \le 1$}
 \STATE {Set inner iteration counter $c=1$,  the convergence parameter $\gamma^{(c)} >  \varepsilon$,  and initialize ${\bf p}^{(c)}$ using the EPA scheme; }
 
 \STATE{Set $\lambda=\lambda+\mu$;}
 \STATE {Calculate $\epsilon=\lambda{R_{\max}}$;}
 \WHILE {$\gamma^{(c)} - \gamma^{(c-1)} > \varepsilon$}
        \STATE {$c = c + 1$;}
        \STATE {Calculate ${\bf y}^{(c)}$ using ${\bf p}^{(c-1)}$ and (\ref{edas1});}
        \STATE {Solve the convex problem (\ref{abt}) given $\epsilon$ and ${\bf y}^{(c)}$ to update ${\bf p}^{(c)}$;}
        \STATE {Set $\gamma^{(c)}$ to the objective function value of (\ref{abt});}
 \ENDWHILE
 \STATE{Calculate   $\eta^{t} = \frac{{R}_T \left({\bf p}^{(c)}, {\bf y}^{(c)}  \right)  } {{P}_T \left({\bf p}^{(c)}, {\bf y}^{(c)}  \right)} $;} 
 \STATE{Set ${\mathcal G}^{t}=\left({\bf p}^{(c)},{\bf y}^{(c)}\right)$};
 \STATE{Update $t=t+1$;}
  \ENDWHILE
 \STATE {{\textbf{Output:}} ${\boldsymbol \eta}$, ${\boldsymbol {\mathcal G}}$}.
 \end{algorithmic}
\end{algorithm}
\setlength{\textfloatsep}{15pt}

In this algorithm, problem (\ref{abt}) is solved repeatedly for the different values of the $\epsilon$ vector. Specifically, by initially setting $\lambda=0$ and increasing it by a small step size $\mu$ such that $\lambda$ always has a positive value in the range of $(0,1]$, different values of $\epsilon$ can be generated according to  $\epsilon=\lambda \times R_{\max}$. The values of $\lambda$ must be in the range of $(0,1]$ because:
\begin{enumerate}
    \item If  $\lambda=0$,  $\epsilon=0$ and, as a result, (\ref{abt}) becomes a transmit power minimization problem without  any constraint on the sum-rate (i.e., $C13$ becomes inactive). Since EE seeks to balance the total power consumed and the achievable rate simultaneously, it becomes imperative to consider the sum rate. Since this is not the case when $\lambda=0$, $\lambda$ must always have a value greater than 0.
    \item If  $\lambda>1$,  $\epsilon>R_{\max}$ and, problem (\ref{abt}) becomes infeasible since, with the given transmit power budgets of the APs and according to (\ref{abn1}), the maximum attainable sum-rate in the aggregated RF/VLC system is $R_{\max}$. Hence, values of $\lambda$ greater than one are not considered.
    \item If $\lambda=1$, $\epsilon=R_{\max}$, and problem (\ref{abt}) becomes a sum-rate maximization problem.
    \item If $0<\lambda<1$,  $0<\epsilon<R_{\max}$ and (\ref{abt}) turns out to be a MOOP.
\end{enumerate}  
Based on the above discussions, values for $\lambda$ in 3) and 4) are used in the proposed algorithm. At the $t$-th iteration of the algorithm, problem (\ref{abt}) is solved for the given values of $\lambda$ and $\epsilon$ to obtain the optimal solution, which is stored as the vector ${\mathcal G}^{t}$ and the corresponding EE solution is denoted as $\eta^t$. Note that ${\mathcal G}$ is the non-dominated set of solutions for the problem in (\ref{eepa}). For any solution outside this set, we can always find a solution in ${\mathcal G}$ that will dominate the former. Thus, ${\mathcal G}$ has the property of dominating all other solutions that do not belong to this set. At the end of the algorithm, the different solutions in ${\mathcal G}$ form the Pareto optimal set. Since it is desired to obtain the best EE performance (i.e., most  appropriate trade-off between sum-rate and total power consumption), the Pareto optimal solutions are ranked based on their corresponding objective function values, and the best is selected as the solution for the EE optimization problem in (\ref{eepa}). This solution is the global optimum since it is determined by solving the convex problems in  (\ref{abn1}) and (\ref{abt}).

\vspace*{-5mm}
\subsection{Complexity of the EE Optimization Solution}
\vspace*{-1.5mm}
The overall complexity of the proposed joint solution for the energy-efficient AP assignment, SA, and transmit PA involves the computations involved in solving each subproblem. For a given SA and PA, the worst-case complexity of the energy-efficient AP assignment algorithm in Algorithm~\ref{A1} can be given as  ${\mathcal O}\left( {\left|{\mathcal{J}} \right| \times S_{\rm AP} \times \log S_{\rm AP}} \right)+{\mathcal O}\left(\left|{\mathcal{J}}\right| \times S_{\rm AP} \right)\approx{\mathcal O}\left( {\left|{\mathcal{J}} \right| \times S_{\rm AP} \times \log S_{\rm AP}} \right)$, where $S_{AP}$ is the number of subchannels for each AP, the term 
${\mathcal O} \left( {\left|{\mathcal{J}} \right| \times S_{\rm AP} \times \log  S_{\rm AP}} \right)$ is the complexity for all users constructing their PLs using off-the-shelf sorting algorithms such as {\textit {merge}} sort and  {\textit {quick}} sort, and the term ${\mathcal O}\left(\left|{\mathcal{J}}\right| \times S_{\rm AP} \right)$  is the complexity of all users proposing to the subchannels of the APs assigned to them. 
For any given AP assignment and PA, the worst-case complexity of the proposed SA procedure for each AP is ${\mathcal O}\left(\left|{\mathcal{K}} \cup {\mathcal{V}} \right| \times  \left|{\mathcal{J}}\right| \times S_{\rm AP} \right)$. In comparison to the optimal SA procedure (i.e., exhaustive search) which has a computational complexity of ${\mathcal O}\left(\left(\left|{\mathcal{K}} \cup {\mathcal{V}} \right|  \right)  S_{\rm AP}!\times 2^{{\mathcal{J}}}   \right)$, the proposed SA scheme has significantly lower complexity.
With regards to the PA subproblem for any given AP assignment and SA, the associated computational complexity comes from solving (\ref{abn1}) and (\ref{abt}). Since (\ref{abn1}) and (\ref{abt}) are convex problems for any given ${\bf y}$, and by following the standard convex analysis in  \cite{boyd}, Algorithms~\ref{A3} and \ref{A4} have a polynomial time complexity in terms of the number of variables (i.e., $\left|{\mathcal{J}}\right| \times S_{AP}$) and constraints (i.e., $\left|{\mathcal{J}}\right|\left(S_{AP}+1\right)+ \left|{\mathcal{K}} \cup {\mathcal{V}} \right|$). From the above complexity analysis, the proposed resource allocation scheme, including AP assignment, SA, and transmit PA, has a polynomial-time worst-case complexity.

\vspace*{-4mm}
\section{Simulation Results}\label{srs}
\vspace*{-0.5mm}
In this section, the performance of the proposed AP assignment, SA, and PA optimization algorithm is investigated in terms of the EE, the sum-rate, and the outage performances of the aggregated RF/VLC system. The macrocell AP is located at the center of the macrocell and has a cell radius of 500 m. The picocell and the VLC APs are randomly and uniformly deployed overlaying the macrocell. Each picocell has a coverage radius of 100 m, and each VLC indoor environment is a room with an area of $5 \times 5$ m$^2$. Each VLC environment has two APs, deployed at the height of 2.15 m, with overlapping coverage to guarantee uniform illumination across the room. The total transmit power for the macrocell AP, each picocell AP, and each VLC AP is 46 dBm, 30 dBm, and 30 dBm, respectively. Each AP has 50 subchannels, with a subchannel bandwidth of 20 MHz and 10 MHz for the VLC and RF systems, respectively. The minimum rate requirement of each user is set as $50$ Mbps. The remaining system model parameters are summarized in Table~\ref{tab1}. The following benchmark schemes and configuration are considered for comparison:

\begin{table*}[t]
\centering
\caption{Simulation Parameters}
\label{tab1}
\begin{tabular}{|ll|ll|}
\hline
\multicolumn{2}{|c|}{\textbf{RF system}}                                                             & \multicolumn{2}{c|}{\textbf{VLC system}}                                                                           \\ \hline
\multicolumn{1}{|l|}{\textbf{Parameter}}                                    & \textbf{Value}         & \multicolumn{1}{l|}{\textbf{Parameter}}                      & \textbf{Value}                                      \\ \hline
\multicolumn{1}{|l|}{Noise power spectral density, $N_{\rm RF}$}                & -174 dBm/Hz            & \multicolumn{1}{l|}{Physical area of PD, $A_{\rm PD}$}           & 1 cm$^2$                                            \\ \hline
\multicolumn{1}{|l|}{Log-normal shadowing standard deviation, $X_{\sigma}$}               & 10 dB                  & \multicolumn{1}{l|}{LED semi-angle at half-power, ${{\phi _{{1}/{2}}}}$}           & $60^{\circ}$                                                  \\ \hline
\multicolumn{1}{|l|}{Multipath fading type}                                 & Rayleigh fading        & \multicolumn{1}{l|}{Gain of the optical filter, $T\left( {{\psi _{v,j}}} \right)$}             & 1                                                   \\ \hline
\multicolumn{1}{|l|}{Circuit power consumption, $P_{\rm PBS}$}                  & 6.8 W \cite{6056691}                 & \multicolumn{1}{l|}{Refractive index, $f$}                   & 1.5                                                 \\ \hline
\multicolumn{1}{|l|}{\multirow{4}{*}{Circuit power consumption, $P_{\rm MBS}$}} & \multirow{4}{*}{130 W  \cite{6056691}} & \multicolumn{1}{l|}{PD responsivity, $R_{\rm PD}$}               & 0.53 A/W                                            \\ \cline{3-4} 
\multicolumn{1}{|c|}{}                                                      &                        & \multicolumn{1}{l|}{FOV of a PD, ${\psi _{{\rm{FoV}}}}$}                             & $70^{\circ}$                                                  \\ \cline{3-4} 
\multicolumn{1}{|c|}{}                                                      &                        & \multicolumn{1}{l|}{Circuit power consumption, $P_{\rm VLC}$}    & 4 W                                                 \\ \cline{3-4} 
\multicolumn{1}{|c|}{}                                                      &                        & \multicolumn{1}{l|}{Noise power spectral density, $N_{\rm VLC}$} & $10^{-21}$ A$^2$/Hz \\ \hline
\end{tabular}
\vspace*{-1mm}
\end{table*}

\begin{itemize}
    \item SCG-SCG-EPA scheme: This scheme assigns APs to users based on the strongest channel power gain (SCG) rule. The available subchannels and  power are allocated according to the SCG rule and the EPA policy, respectively, for the users assigned to an AP.
    \item Baseline scheme: This scheme has been adopted from \cite{8314706}, in which the authors proposed an SA and PA procedure for an aggregated RF/VLC system under the assumption that users are assigned to the AP with the SCG. Specifically, the AP assignment scheme is according to the SCG rule, the SA scheme is according to Algorithm~2 of \cite{8314706}, and the PA scheme is per our proposed energy-efficient PA scheme. 
    \item Hybrid RF/VLC: In this configuration, each user is only  assigned a macrocell AP or a picocell AP or a VLC AP (i.e., $\sum\limits_{\forall k} x_{k,j} + \sum\limits_{\forall v} x_{v,j}=1,\,\forall j$). The proposed iterative solution for the joint problem is adopted for this configuration.
\end{itemize}

\begin{figure}[!t]
 \centering
\begin{minipage}{0.48\textwidth}
 \centering
 \includegraphics[width=1.1\linewidth]{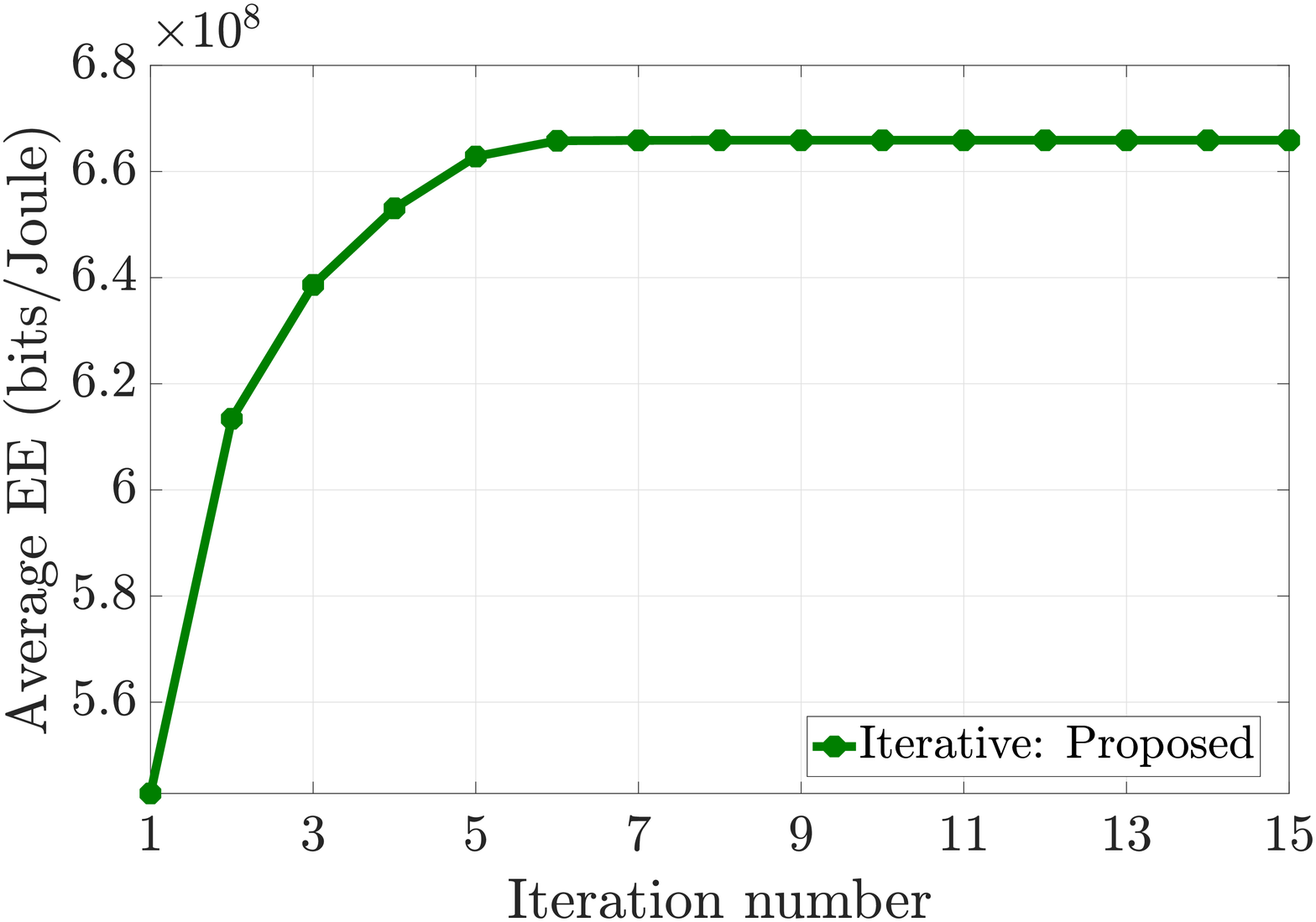}
      \caption{Convergence of the proposed iterative joint solution.}
       \label{res0}
\end{minipage}%
 \hfill
 \begin{minipage}{0.48\textwidth}
 \centering
 \includegraphics[width=1.1\linewidth]{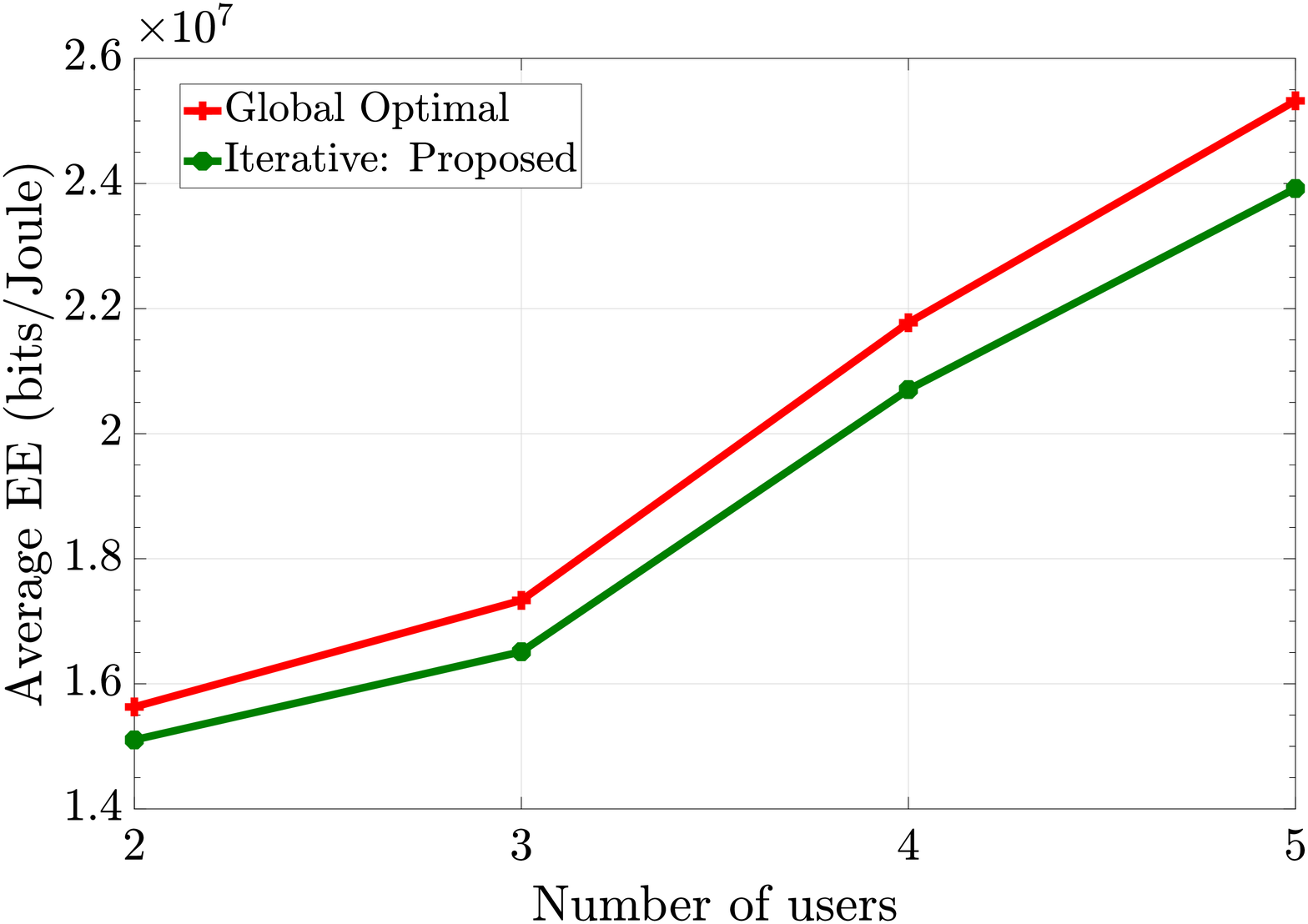}
      \caption{EE comparison of the proposed and the global optimal scheme.}
       \label{reexs1}
\end{minipage}
\vspace*{-2mm}
\end{figure}

Figure~\ref{res0} shows the convergence of the proposed joint solution as demonstrated in Fig.~\ref{solf}(a). It can be seen that the proposed approach converges to a stationary point after 6 iterations.

Figure~\ref{reexs1} illustrates the EE performance gap between the proposed scheme and the globally optimal solution obtained via exhaustive search. As can be seen, the average gap between the proposed and the exhaustive scheme is around $5\%$. This indicates that the proposed sub-optimal scheme approaches the globally optimal solution while offering a practical solution to the joint optimization problem of AP assignment, SA, and transmit PA.     

\begin{figure}[!t]
 \centering
\begin{minipage}{0.48\textwidth}
 \centering
 \includegraphics[width=1.1\linewidth]{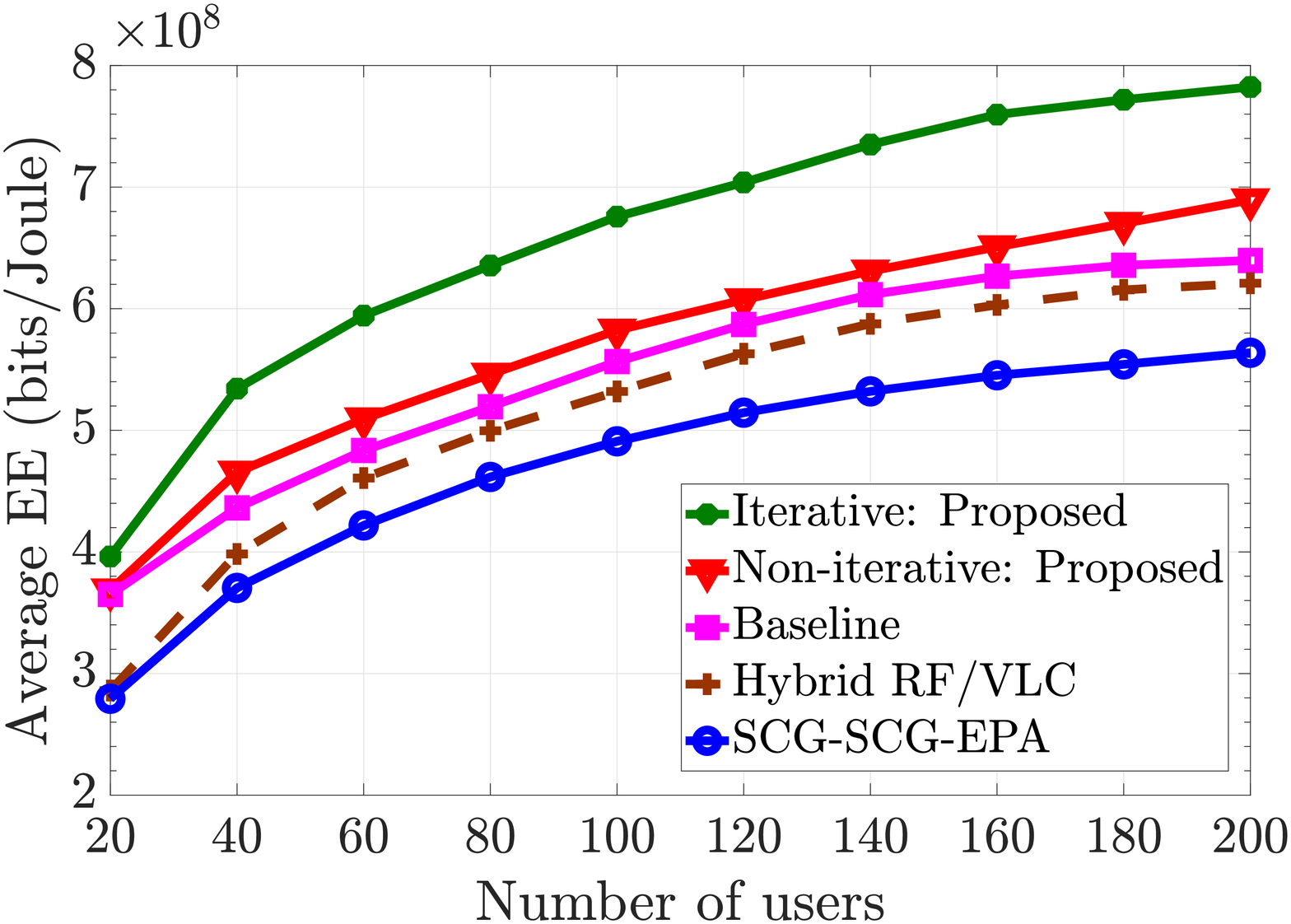}
      \caption{Average EE versus the total number of users.}
       \label{res1}
\end{minipage}%
 \hfill
 \begin{minipage}{0.48\textwidth}
 \centering
 \includegraphics[width=1.1\linewidth]{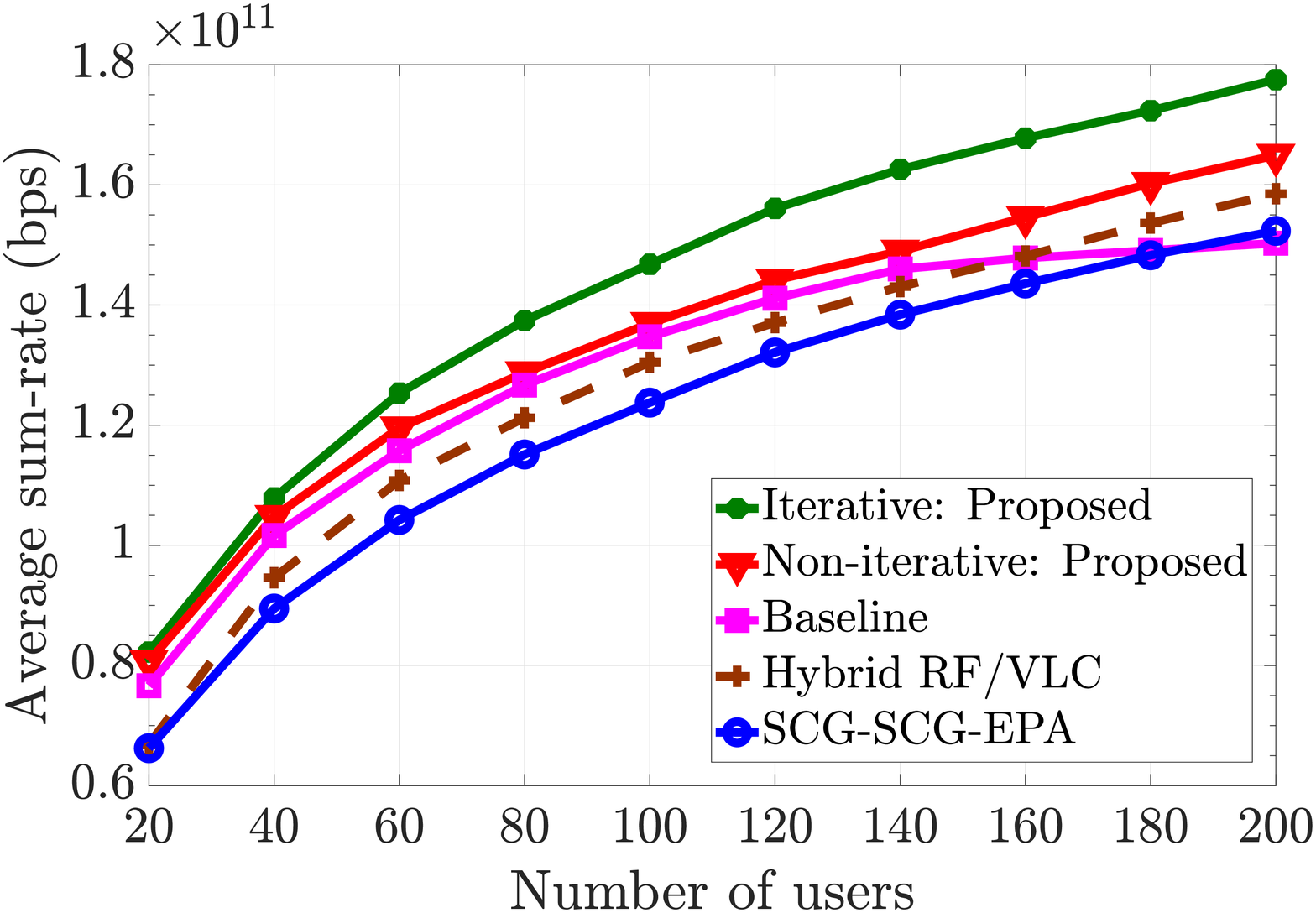}
      \caption{Average sum-rate versus the total number of users.}
       \label{res2}
\end{minipage}
\vspace*{-2mm}
\end{figure}

Figure~\ref{res1} illustrates the average EE performance of the proposed iterative and non-iterative energy-efficient resource allocation schemes, the considered benchmarks, and the hybrid system for varying numbers of users. It can be seen that the proposed schemes outperform the two benchmarks and the hybrid RF/VLC system, and the EE performance improves as the number of users increases for all schemes. This is because the proposed approaches make use of the EE's definition and consider ICI effects when assigning APs to users and during the optimization of the transmit APs' transmit power to users. The energy-efficient MT-based AP assignment scheme in Algorithm~\ref{A1} can assign APs to users under a given PA, such that the overall best EE performance is achieved through the use of preference relations among users and APs. Once the users have been associated with the APs that guarantee the highest EE, the low-complexity SCG rule-based SA procedure efficiently allocates any subchannel of an AP to the user with the best channel condition. After the SA step, the proposed energy-efficient PA scheme is used to update the transmit power to the users. Thus, both the AP assignment and PA schemes of our proposed solution to the EE optimization problem consider the objective function definition and the ICI effects in their decision-making processes. In contrast, the baseline approach does not consider users' EE performance when assigning APs since it uses the SCG rule to assign APs to users. Moreover, the SA policy of the baseline approach focuses more on ensuring that the required minimum rate is guaranteed for all users since it first allocates subchannels to users to guarantee their QoS requirements. 
Note that this results in a trade-off between assigning subchannels to users with the best channel condition and assigning subchannels to improve fairness among users. Specifically, the baseline scheme can assign a subchannel to a user with a relatively worse channel condition to meet the QoS requirement. This can cause the AP to transmit at a higher power level on that subchannel and thus generate significant interference to nearby users being served on that same subchannel. The SCG-SCG-EPA scheme performs worst since it does not consider the definition of EE and ICI effects when assigning APs to users, allocating subchannels, and allocating transmit power. Between the two joint solution approaches, it can be observed that alternating among the  AP assignment, SA, and PA subproblems (i.e., the iterative approach) results in a superior EE performance when compared with the non-iterative approach. However, this performance improvement is obtained at the expense of an additional number of iterations. Moreover, the rate of increase in the EE decreases for all schemes with an increasing number of users due to the resulting stronger impact of ICI. The hybrid RF/VLC system performs worse than the aggregated system (except with the SCG-SCG-EPA scheme). This is because there are more available options for the aggregated system (i.e., there is better exploitation of the available resources) as the users can receive data transmission from both RF and VLC APs. 

Figure~\ref{res2} shows the average sum-rate performance of the proposed schemes, the two benchmarks, and the hybrid RF/VLC system for a varying number of users. Clearly, the proposed schemes outperform the two benchmarks in terms of the average sum-rate, revealing the benefit of jointly optimizing the AP assignment, SA, and PA. Moreover, the sum-rate performance improves with increasing number of users for all four schemes. This is because increasing the number of users for a fixed number of subchannels and transmit power budget expands the feasible region of the considered optimization problem. However, the rate of increase for the baseline scheme decreases after a total of 140 users. This behavior is due to the SA policy used by the baseline scheme. More precisely, this policy fails to exploit all users' channel power gain differences, especially for a more significant number of users, since it always focuses on satisfying the QoS requirements of all users first. 
The performance of the aggregated system (with the proposed schemes) is significantly better than the hybrid system because of the multi-homing capability of the users' receiving devices. Specifically, any user served simultaneously by an RF AP and a VLC AP can have communication links with better channel conditions with at least one AP. This can lead to an improvement in the achieved data rate. Since the proposed iterative approach outperforms the non-iterative approach and the hybrid RF/VLC system, the proposed iterative approach is considered in this paper's remaining EE performance analyses. 

\begin{figure}[!t]
 \centering
\begin{minipage}{0.48\textwidth}
 \centering
  \includegraphics[width=1.1\linewidth]{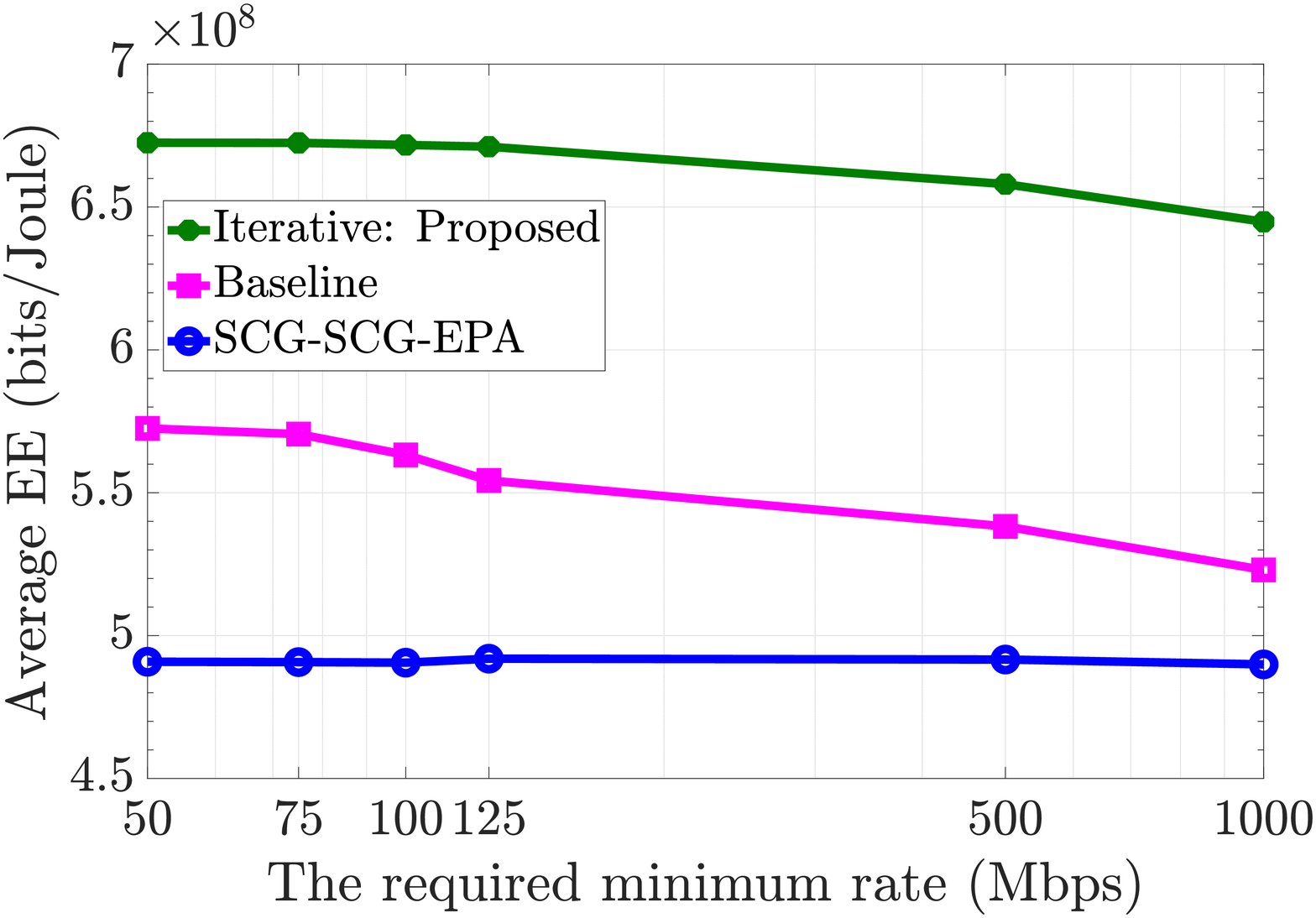}
      \caption{Average EE versus the required minimum rate.}
       \label{res5}
\end{minipage}%
 \hfill
 \begin{minipage}{0.48\textwidth}
 \centering
 \includegraphics[width=1.1\linewidth]{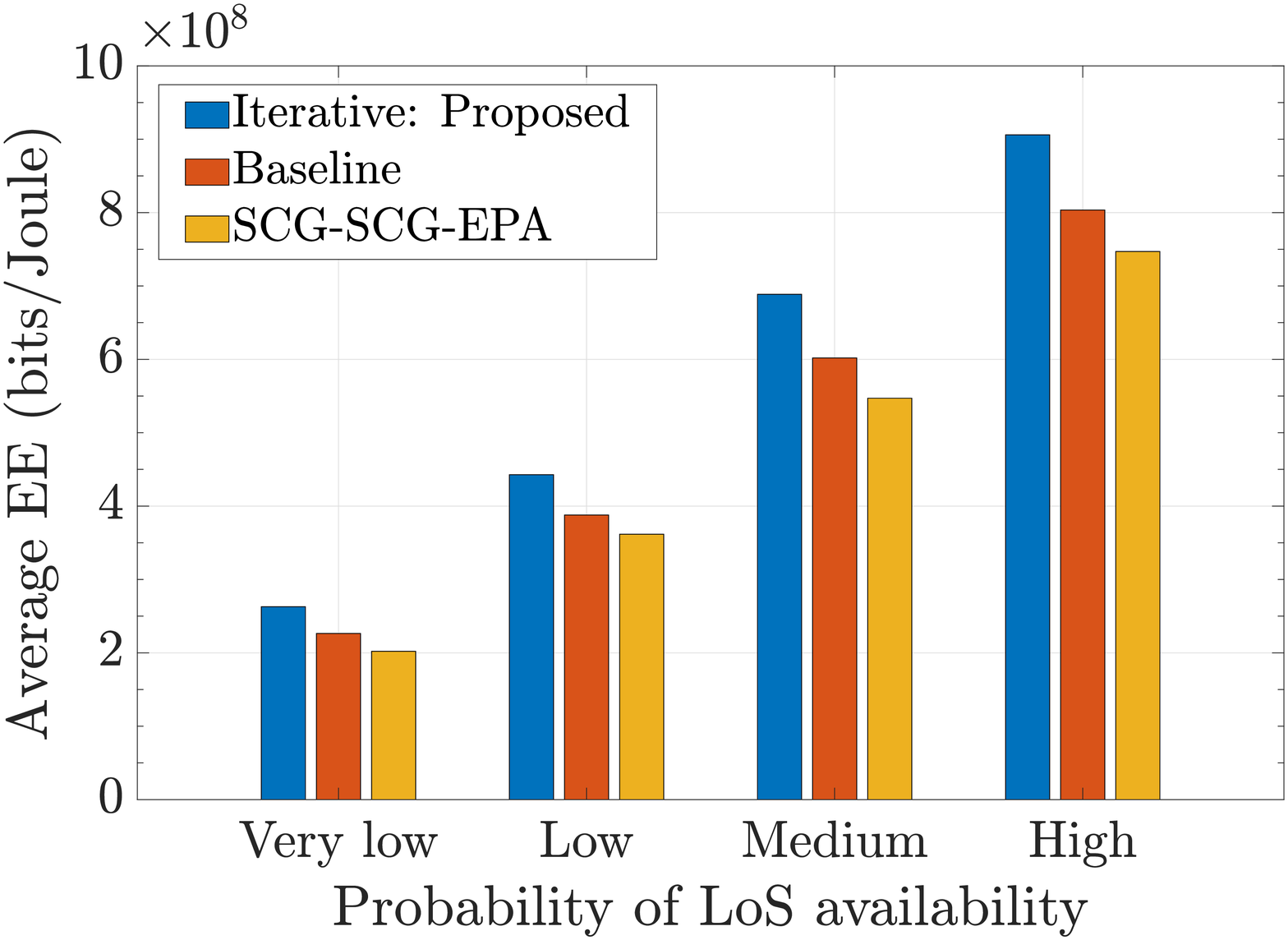}
      \caption{Average EE for different values of the probability of LoS communication.}
       \label{res6}
\end{minipage}
\vspace*{-2mm}
\end{figure}  

Figure~\ref{res5} depicts the average EE of the aggregated RF/VLC system versus the required minimum rate, which is varied from 50 Mbps to 1 Gbps. It can be observed that the average EE decreases as the required minimum rate value increases for both the proposed and the baseline schemes. This can be explained by the fact that increasing the value of $R_{\min}$ restricts the feasible region of the EE optimization problem. However, the trend grows much slower (especially between 50 to 125 Mbps) for the proposed solution compared to the baseline approach. This indicates the ability of our proposed approach to cope very well with higher users' rate requirements. On the contrary, increasing the minimum rate requirements of users does not affect the EE curve for the SCG-SCG-EPA scheme since the value of $R_{\min}$ is never used in the assignment of APs or the allocation of power  and subchannel resources by this naive scheme. 
 
Figure~\ref{res6} demonstrates how the existence of LoS communication links between VLC APs and users influences the EE performance of the proposed solution and the two benchmarks. The labels on the x-axis of this figure are defined as follows: ``Very low'' indicates that the probability of an LoS path is  between 0 and 0.3; ``Low'' means the probability is between 0.3 and 0.5; ``Medium'' means the probability is between 0.5 and 0.8; ``High'' means the probability is between 0.8 and 1. Note that such labels allow the probability of an LoS scenario to vary among the various users and the access points, which depicts the fact that the VLC channel changes with changing the location of the receivers. The figure reveals that LoS path blockage affects the EE performance of all the considered schemes as it considerably impacts the propagation environment. Specifically, when the probability of having an LoS path is low, users are less likely to be served by the AP that can provide the highest EE. This would lead to a reduction in the overall EE. The proposed scheme has the best EE performance for the different scenarios considered.   

\begin{figure}[!t]
 \centering
\begin{minipage}{0.48\textwidth}
 \centering
\includegraphics[width=1.1\linewidth]{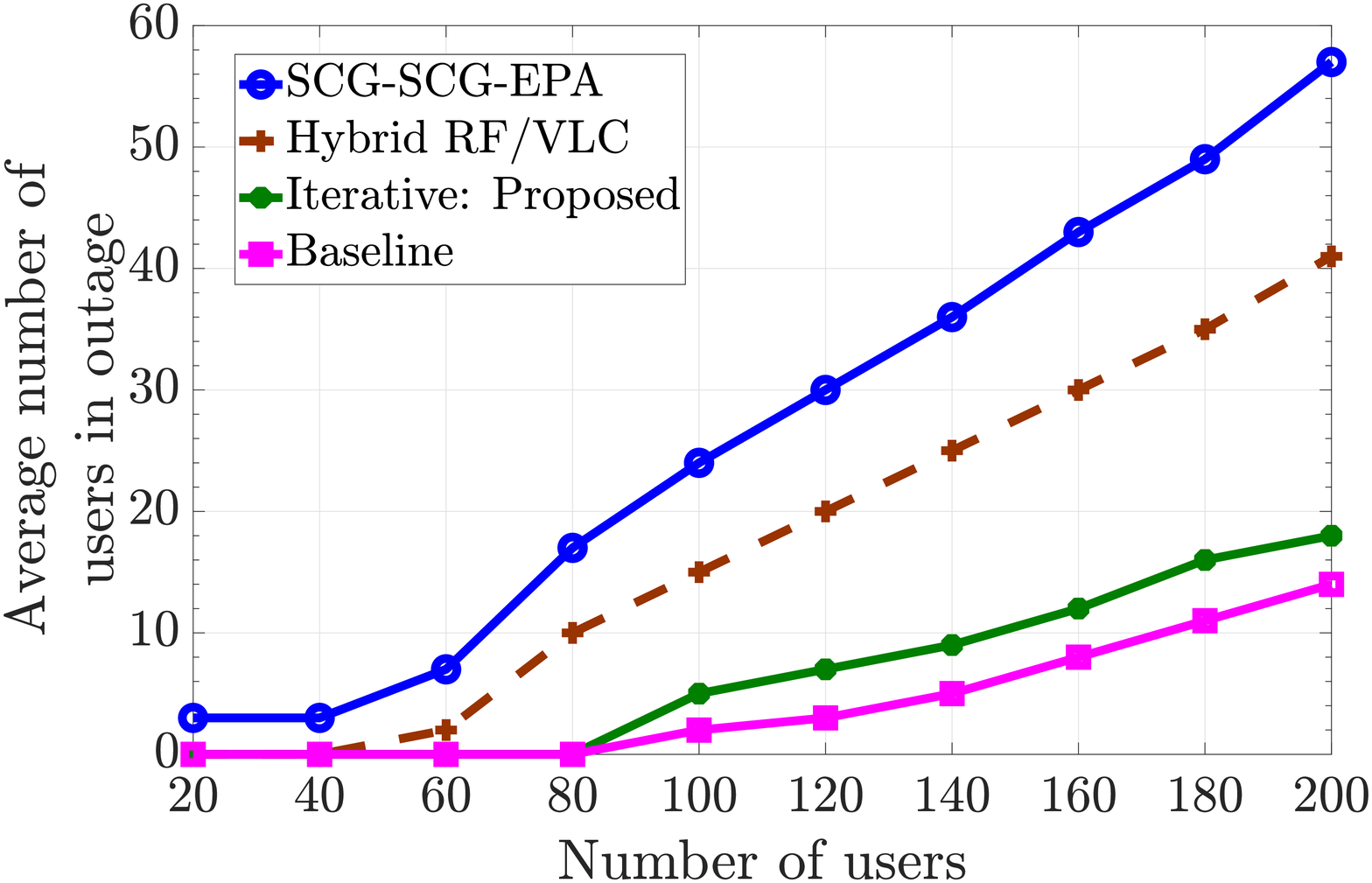}
      \caption{Average number of users in outage versus the total number of users.}
       \label{res3}
\end{minipage}%
 \hfill
 \begin{minipage}{0.48\textwidth}
 \centering
 \includegraphics[width=1.1\linewidth]{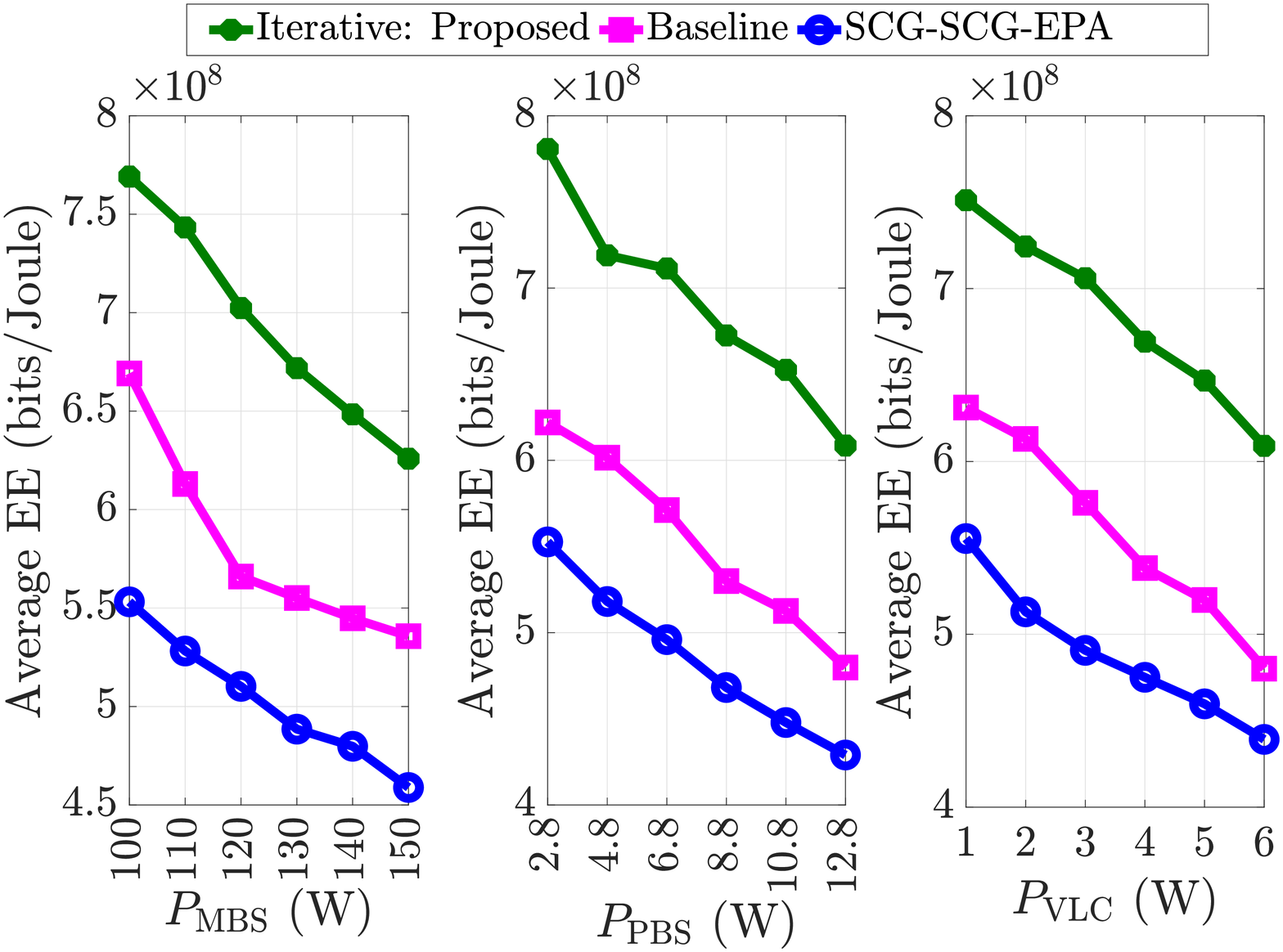}
      \caption{Average EE versus the circuit power consumption.}
       \label{res4}
\end{minipage}
\vspace*{-2mm}
\end{figure}  

Figure~\ref{res3} shows the average number of users whose QoS requirements cannot be guaranteed by the three schemes versus the total number of users. The SCG-SCG-EPA scheme has the highest number of users in outage since this scheme does not take 
into consideration users' minimum rate requirements when allocating the available resources. The hybrid system performs worst compared with the proposed and baseline schemes since users with relatively bad channel conditions and assigned to an AP may not be allocated any subchannel. As a result, their QoS requirements cannot be guaranteed. However, this is not the case in the aggregated system since users can receive from both RF and VLC APs to realize additional data rates to meet their QoS requirements. The baseline scheme outperforms the proposed approach since it prioritizes satisfying the minimum rate requirement of all users first by allocating subchannel resources to them such that the $R_{\min}$ value is achieved for most or all users. Moreover, the outage increases with increasing users for all schemes due to increased competition for the limited resources.

Figure~\ref{res4} indicates the average EE performance of the proposed scheme, the baseline scheme, and the SCG-SCG-EPA scheme when the values of the circuit power consumption for the macrocell, picocell, and VLC APs are varied. Note that the range of the considered circuit power values lies within the typical practical ranges for MBSs \cite{6056691}, PBSs \cite{6056691}, and VLC APs \cite{7437374}. It can be observed that the EE decreases for all the schemes as the circuit power consumption increases for the macrocell, picocell, and VLC APs. This observation is in line with the definition of the system EE. Hence, future designs should focus on hardware components with lower circuit power consumption. 

\vspace*{-4mm}
\section{Conclusion}\label{ccs}
This paper has investigated EE optimization for aggregated RF/VLC systems by jointly optimizing AP assignment, SA, and PA. More specifically, the original EE optimization problem, which belongs to the class of mixed-integer nonlinear programming problems and is generally intractable, has been decoupled into AP assignment, SA, and transmit PA subproblems. A solution technique has been proposed for each, and two frameworks to obtain the joint solution have been introduced. A novel energy-efficient AP assignment scheme has been developed by invoking MT. Additionally, a simple yet efficient SA scheme has been designed based on the AP assignment result. Given the AP assignment and SA solutions, a PA algorithm based on the quadratic transform approach and from the viewpoint of multi-objective optimization has been proposed. Simulation results have  demonstrated the effectiveness of the proposed algorithms. They have revealed the superior EE and sum-rate performances of aggregated RF/VLC systems compared with hybrid systems and existing schemes. Moreover, the impact of critical system parameters, such as the circuit power consumption, users' QoS requirements, and LoS availability for the VLC links, on the performance of the aggregated RF/VLC system has been examined. This paper has revealed that the considered aggregated system and the proposed algorithms can effectively combine resources from both RF and VLC APs to enhance the EE, sum-rate, and outage performances while offering ubiquitous connectivity solution to users. Interesting problems for future work in this area include developing techniques for supporting mobile users and for VLC LoS blockage-aware resource allocation in aggregated systems. 
\bibliographystyle{IEEEtran}
\bibliography{aggregate_draft}
\end{document}